\pdfoutput=1
\documentclass[journal,twoside,web]{ieeecolor}
\usepackage{generic}
\usepackage{cite}
\usepackage{amsmath,amssymb,amsfonts}
\usepackage{graphicx}
\usepackage{hyperref}
\hypersetup{hidelinks=true}
\usepackage{textcomp}
\usepackage{multirow}
\usepackage{mathrsfs}
\usepackage{xcolor} 
\usepackage{manyfoot}
\usepackage{booktabs}
\usepackage{algorithmicx}
\usepackage{algpseudocode}
\usepackage{listings}
\usepackage{tabularx}
\usepackage{subcaption}  
\usepackage{lineno,hyperref}
\usepackage{array} 
\usepackage{colortbl}

\def\BibTeX{{\rm B\kern-.05em{\sc i\kern-.025em b}\kern-.08em
    T\kern-.1667em\lower.7ex\hbox{E}\kern-.125emX}}
\begin{document}
\title{ASP-VMUNet: Atrous Shifted Parallel Vision Mamba U-Net for Skin Lesion Segmentation}

\author{Muyi Bao, Shuchang Lyu, Zhaoyang Xu, Qi Zhao, Changyu Zeng, Wenpei Bai, Guangliang Cheng
\thanks{
Muyi Bao and Changyu Zeng is with the School of Advanced Technology, Xi’an Jiaotong-Liverpool University (email: MuyiBao21@student.xjtlu.edu.cn) \\
Shuchang Lyu and Qi Zhao are with School of Electronics and Information Engineering, Beihang University. \\
Zhaoyang Xu is with the Department of Paediatrics, Cambridge University. \\
Wenpei Bai is with the Department of Gynecology and Obstetrics, Beijing Shijitan Hospital, Capital Medical University. \\
Guangliang Cheng is with the Department of Computer Science, University of Liverpool. \\
Corresponding Author: Guangliang Cheng, Wenpei Bai
}
}

\maketitle

\begin{abstract}
Skin lesion segmentation is a critical challenge in computer vision, and it is essential to separate pathological features from healthy skin for diagnostics accurately. Traditional Convolutional Neural Networks (CNNs) are limited by narrow receptive fields, and Transformers face significant computational burdens. This paper presents a novel skin lesion segmentation framework, the \textbf{A}trous \textbf{S}hifted \textbf{P}arallel \textbf{V}ision \textbf{M}amba \textbf{UNet} (\textbf{ASP-VMUNet}), which integrates the efficient and scalable Mamba architecture to overcome limitations in traditional CNNs and computationally demanding Transformers. The framework introduces an atrous scan technique that minimizes background interference and expands the receptive field, enhancing Mamba's scanning capabilities. Additionally, the inclusion of a Parallel Vision Mamba (PVM) layer and a shift round operation optimizes feature segmentation and fosters rich inter-segment information exchange. A supplementary CNN branch with a Selective-Kernel (SK) Block further refines the segmentation by blending local and global contextual information. Tested on four benchmark datasets (ISIC16/17/18 and PH2), ASP-VMUNet demonstrates superior performance in skin lesion segmentation, validated by comprehensive ablation studies. This approach not only advances medical image segmentation but also highlights the benefits of hybrid architectures in medical imaging technology. Our code is available at \href{https://github.com/BaoBao0926/ASP-VMUNet/tree/main}{https://github.com/BaoBao0926/ASP-VMUNet/tree/main}.
\end{abstract}


\section{Introduction} \label{sec:Introduction}


Deep learning technique \cite{unet} is revolutionizing the healthcare industry, offering innovative solutions for skin lesions. These skin diseases pose a significant public health challenge, with early detection being crucial for improving survival rates \cite{skinSurvey}. Segmentation is key in precisely identifying lesions from healthy skin \cite{UMamba}. U-Net \cite{unet}, a popular architecture for medical image segmentation, has been proven effective in this task, owing to its encoder-decoder structure with skip connections that enhance feature fusion. 
Recently, Convolutional Neural Network (CNN)-based U-Net \cite{unet} and Transformer-based U-Net \cite{TransUnet, UNETR} have shown exceptional performance in segmentation tasks.

Despite the significant success achieved by both CNN-based and Transformer-based models, inherent limitations hinder their advancement. CNN-based models are constrained by their local receptive field, which significantly limits their capacity to capture long-range dependencies and leads to suboptimal performance. On the other hand, Transformer-based models excel at global modeling but face the quadratic complexity of the self-attention mechanism to image size \cite{transformer, VIT}. This high computational cost is unacceptable in dense prediction tasks. 

To address these challenges posed by CNNs and Transformers, Mamba \cite{mamba} offers a promising alternative, which can effectively capture long-range dependencies while maintaining linear computational complexity regarding the image size. Originally developed for Natural Language Processing (NLP), Mamba-based models have been introduced into the computer vision field \cite{VisionMamba, VMamba, VMUNet, VMUNetv2}. In image processing, the common practice is to divide an image into smaller patches and then need to be scanned (transformed) into 1D sequences that are input into the Mamba. Several studies have proposed various scanning methods to enhance its 2D spatial perception in 1D sequences \cite{VisionMamba, VMamba, LocalMamba, plainMamba, efficientScan, sliceMamba}. However, these scan methods are unsuitable for skin disease images. We observe that there are many background patches with limited information, which may disrupt the hidden state and cause the model to lose previously acquired information, thereby degrading its performance.

\begin{figure*}[ht]  
    \centering
    \includegraphics[width=0.95\textwidth]{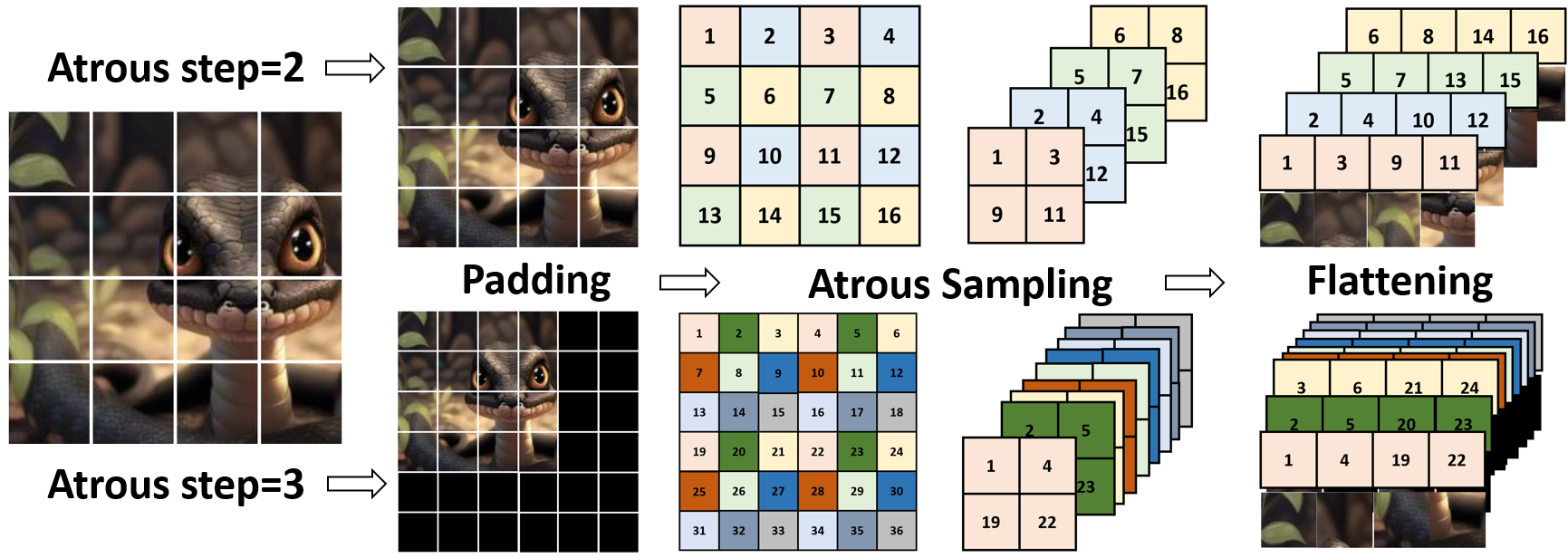}  
    \caption{Atrous scan process: First, pad the image to ensure the height and width are divisible by the atrous step. Then, sample the image into sub-images. Finally, flatten the sub-images into a 1D sequence for input into the Mamba block.}
    \label{fig:Atrous Scan}
\end{figure*}

Motivated by this observation and atrous CNNs \cite{atrousCNN}, we propose a novel scanning method, called \textit{atrous scan}. The process is illustrated in Fig. \ref{fig:Atrous Scan}. Let $H$ and $W$ represent the height and width of the image in terms of patches. First, padding is applied to ensure that both $H$ and $W$ are divisible by the atrous step $S$. The image is then sampled at intervals of $S$ to generate $S^2$ sub-images. These sub-images are subsequently flattened into 1D sequences and then are input into Mamba. Unlike conventional downsampling techniques such as max-pooling, our method does not discard any patches, while it also can reduce the number of less informative patches within each 1D sequence and simultaneously increase the receptive field. This approach not only improves the performance but also allows for the seamless integration of other scanning methods as a plug-and-play module, as described in Section \ref{sec:category}.

Furthermore, we employ the Parallel Vision Mamba (PVM) layer \cite{ultralightVMUnet} as the main Mamba block, which divides the input features into multiple segments (four in this paper) along the $C$-dimension. These feature segments are processed in parallel by one Mamba, thus reducing the number of parameters by 74.8\% while maintaining competitive performance. However, inter-segment information is less exchanged since four feature segments are inputted into the Mamba block independently. Therefore, in order to address this problem, inspired by the Swin-Transformer \cite{SwinTransformer}, we introduce the $shift$ $round$ operation, which recombines the feature segments between neighboring segments. This operation facilitates interaction between segments and strengthens the overall feature representation capacity, thereby improving the model performance.

It is worth noting that regional representation is useful for detailed pathological detection. Following \cite{efficientScan}, we design a CNN branch to supplement the global information extracted by Mamba with local information. However, we note that existing parallel CNN-Mamba hybrid models \cite{efficientScan, DeformMamba} simply add the global and local features together, leading to suboptimal feature fusion. For example, local features are crucial at the boundary between skin lesions and healthy skin, since the CNN's local receptive field can better capture edge details. To mitigate this problem, we introduce SK Block \cite{SKNet}, which uses the attention mechanism to effectively fuse local and global information, ensuring a more robust feature representation.

In this paper, we propose a novel Mamba-based U-Net model, namely \textbf{A}trous \textbf{S}hifted \textbf{P}arallel \textbf{V}ision \textbf{M}amba \textbf{UNet} (\textbf{ASP-VMUNet}), which incorporates the \textit{atrous scan}, \textit{shift round} operation and SK Block for more accurate skin lesion segmentation. In summary, our paper's contributions are as follows:

\begin{itemize}
\item We introduce a novel scanning method, $atrous$ $scan$, which reduces the impact of less informative background patches and increases the receptive field, enhancing the model's ability to preserve relevant information. This method, as a plug-and-play module, can be easily integrated with existing scanning approaches.
\item We propose the $shift$ $round$ operation to improve the information exchange between feature segments. Additionally, we introduce the SK Block, which leverages the attention mechanism to effectively fuse global and local features, avoiding information loss caused by direct addition.
\item Through extensive experiments on four widely used skin lesion segmentation datasets (ISIC16/17/18 and PH2), we demonstrate that our model achieves state-of-the-art (SOTA) performance. A comprehensive ablation study further validates the contribution of each component, the rationality of model configuration, and the effectiveness of the atrous scan.
\end{itemize}

\section{Related Work}
\subsection{Vision Mamba Backbone}
Since the birth of Mamba, numerous studies have investigated its performance in the field of CV and proposed various vision backbone models. Vision Mamba (Vim) \cite{VisionMamba} was the first to utilize Mamba as a fundamental block in CV, incorporating class tokens into image patch tokens and introducing a novel bidirectional scan method. This method scans the image in both forward and backward directions, thereby enhancing the model's 2D spatial representation capabilities. Visual Mamba (VMamba) \cite{VMamba} introduced the VSS block, which employed an across scan. This method scans the entire image from four directions to enhance 2D spatial awareness. In addition to these approaches, other Mamba-based vision backbones have been developed, such as Local Mamba \cite{LocalMamba}, Efficient Mamba \cite{efficientScan}, and Mamba-ND \cite{MambaND}, which have demonstrated superior performance compared with both CNN-based and Transformer-based models.

\subsection{Mamba-Based U-Net In Medical Segmentation}
Inspired by the success of the vision Mamba backbone model, some studies have explored the Mamba-based U-Net framework for medical image segmentation tasks. U-Mamba \cite{UMamba} was the first to incorporate Mamba into the U-Net architecture, designing a hybrid CNN-Mamba block to enhance the model’s ability to capture long-range dependencies. SegMamba \cite{SegMamba} introduced a Gated Spatial Convolution (GSC) block to strengthen spatial awareness, followed by a Tri-orientated Spatial Mamba, initially designed for 3D data, to further enhance segmentation performance. VM-UNet \cite{VMUNet} is the first pure Mamba-based model for segmentation, utilizing the VSS block as the primary component in both the encoder and decoder. LightM-UNet \cite{lightmunet} is the first work to focus on a lightweight model. Compared with the 173M parameters of U-Mamba, LightM-UNet significantly reduces the parameter count to just 1.87M. Furthermore, UltraLight-VMUNet \cite{ultralightVMUnet} introduces the PVM layer, which divides the features along the $C$-dimension, further reducing the parameter count to 0.05M. In addition to these works, other studies have explored different potentials of Mamba, such as T-Mamba \cite{TMamba}, which utilizes the frequency domain to enhance feature extraction in the context of tooth segmentation, and VMUNet-v2 \cite{VMUNetv2}, which firstly modifies the skip connection mechanism to improve the fusion of low-level and high-level features, 
In this paper, we propose a parallel CNN-Mamba hybrid U-Net model that incorporates with atrous scan, SE module and SK module, which achieves SOTA on four common-used skin lesion datasets.

\subsection{Scan Method}
As described in Section \ref{sec:Introduction}, the scan method refers to the process of transforming an image into one or more 1D sequences, which constitutes the core component of Mamba. In addition to the bidirectional scan in Vim \cite{VisionMamba} and the across scan in VMamba \cite{VMamba}, numerous studies have proposed alternative scan methods. 
Local scan in Local Mamba \cite{LocalMamba} introduces local scan that scans area by area (2 $\times$ 2 or 7 $\times$ 7), which can enhance the ability to focus on an area. Plain Mamba \cite{plainMamba} introduces the plain scan, which enhances spatial continuity by ensuring the adjacency of tokens within the scanning sequence and starting from the four corner patches both vertically and horizontally. Slice Mamba \cite{sliceMamba} proposes the slice scan, which guarantees that spatially adjacent features remain close within the scanning sequence. Efficient Mamba \cite{efficientScan} introduces the efficient scan, which leverages the concept of atrous to reduce the computational burden associated with across scans. Other works, such as SegMamba \cite{SegMamba} and Video Mamba \cite{videomamba}, extend scan methods to 3D data. 
In this paper, we propose \textit{atrous scan}, whose core idea is to minimize the interference of less informative background patches and increase the receptive field. Additionally, we categorize these scan methods into three distinct categories: scan mode, scan direction, and scan sampling, as detailed in Section \ref{sec:category}. To the best of our knowledge, \textit{atrous scan} is the first scan sampling method besides global scan sampling, which can be incorporated into any existing scan methods mentioned above. 

\subsection{Mamba-CNN Parallel Network}
Several studies have designed two parallel branches, i.e. one CNN branch and one Mamba branch, within one basic block. This aims to supplement global features with local features extracted by CNN. Efficient Mamba \cite{efficientScan} employed a depth-wise CNN branch, while Deform-Mamba \cite{DeformMamba} designed a deformable CNN branch \cite{deformableCNN}. Both of them directly conduct element-wise addition of the features from the two branches. In contrast, our approach mitigates potential information loss caused by direct addition by introducing the SK Block \cite{SKNet}, which facilitates more effective feature fusion.

\section{Methodology} \label{Methodology}

\subsection{Architecture Overview} \label{archOverview}

\begin{figure*}[ht]  
    \centering
    \includegraphics[width=0.95\textwidth]{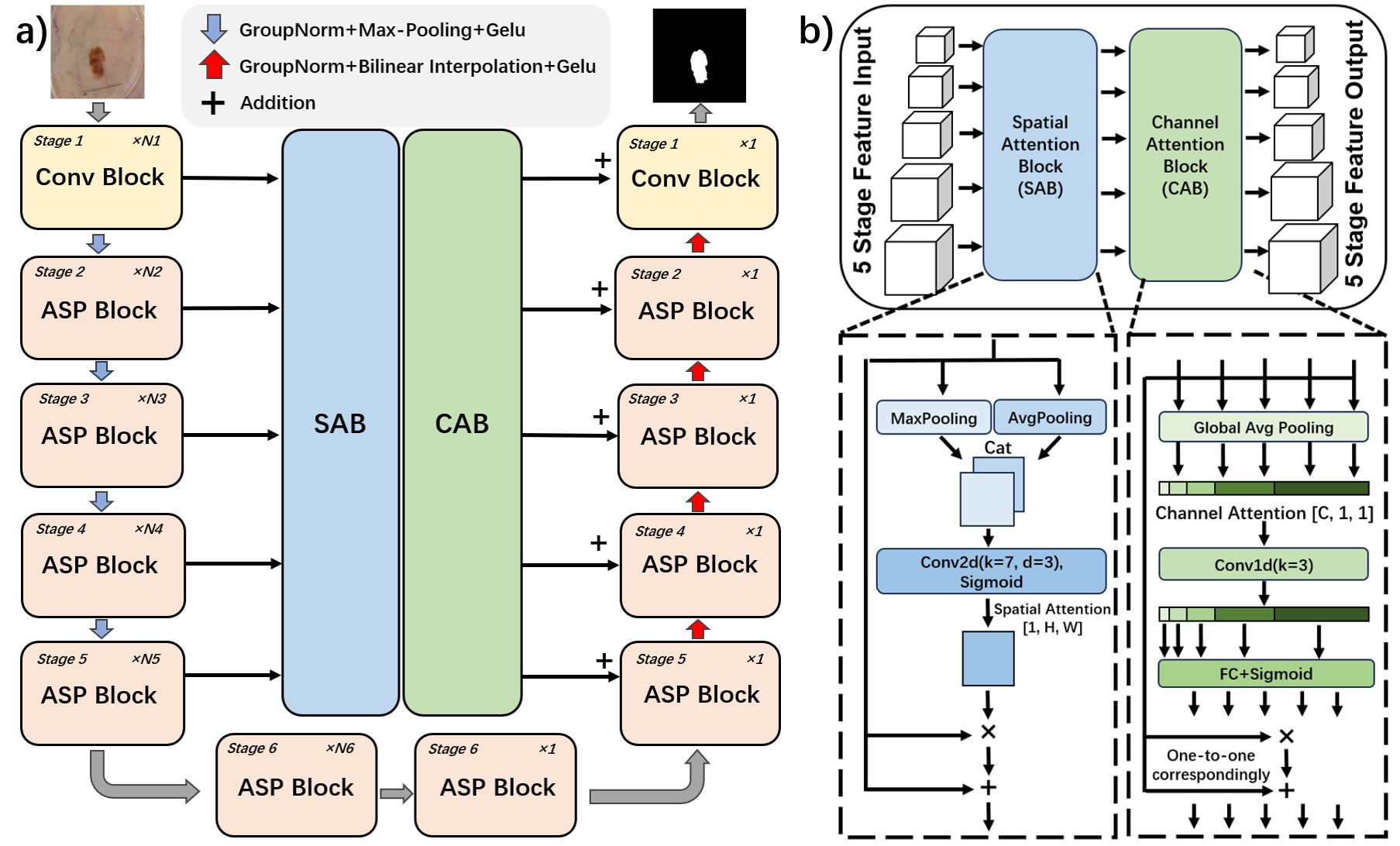}  
    \caption{ 
    (a) shows an overview of ASP-VMUNet architecture: the network follows the U-Net framework with 6 stages in both encoder and decoder. The first stage is the CNN block and the subsequent is the ASP block in both the encoder and decoder. Skip connections incorporate shared-weighted Spatial Attention Blocks (SAB) and Channel Attention Blocks (CAB). (b) shows the process of SAB and CAB block, utilizing an attention mechanism to fuse multi-level and multi-scale features.
    }
    \label{fig:architecture overview}
\end{figure*}

The overall architecture of ASP-VMUNet is illustrated in Fig. \ref{fig:architecture overview}. Following the U-Net \cite{unet} framework, ASP-VMUNet comprises a total of 6 stages in both the encoder and decoder. The first stage employs the CNN block, while the subsequent stages employ the Atrous Shifted Parallel (ASP) block, which is described in Section \ref{ASPBlock}. The configuration of the encoder's layers is defined as \textit{[N1, N2, N3, N4, N5, N6]}, with the decoder uniformly comprising one layer. The skip connection uses one shared-weighted Spatial Attention Block (SAB) and one Channel Attention Block (CAB), which fuse multi-level and multi-scale information from the first five stages \cite{malunet}.

\subsection{Spatial and Channel Attention Block} \label{SCAB}

We introduce the Spatial Attention Block (SAB) and Channel Attention Block (CAB) \cite{malunet,CBAM} to fuse multi-stage and multi-scale information, addressing the challenges posed by varying image sizes and the potential for redundant information in features. After processing through the SAB and CAB, features at each stage of skipping connection can pay more attention to important areas in the image. The SAB and CAB are shown in Fig. \ref{fig:architecture overview}b.

SAB first concatenates the max-pooled and average-pooled feature maps, which are subsequently passed through a CNN layer with a kernel size of 7 and dilation rate of 3, and a sigmoid activation function to generate an attention map. The output is then computed by element-wise multiplication of the input with the attention map and then addition with the original input.
CAB concatenates the feature maps of five different stages that apply global average pooling (GAP). The feature maps are passed through a 1D CNN layer with a kernel size of 3, a linear layer and a sigmoid activation function to generate the attention map. Finally, the output is obtained by element-wise multiplication of the input feature with its corresponding attention map and then addition back to the original input.





\subsection{Atrous Shifted Parallel Block} \label{ASPBlock}
\subsubsection{Overview of ASP Block}
\begin{figure*}[ht]  
    \centering
    \includegraphics[width=0.95\textwidth]{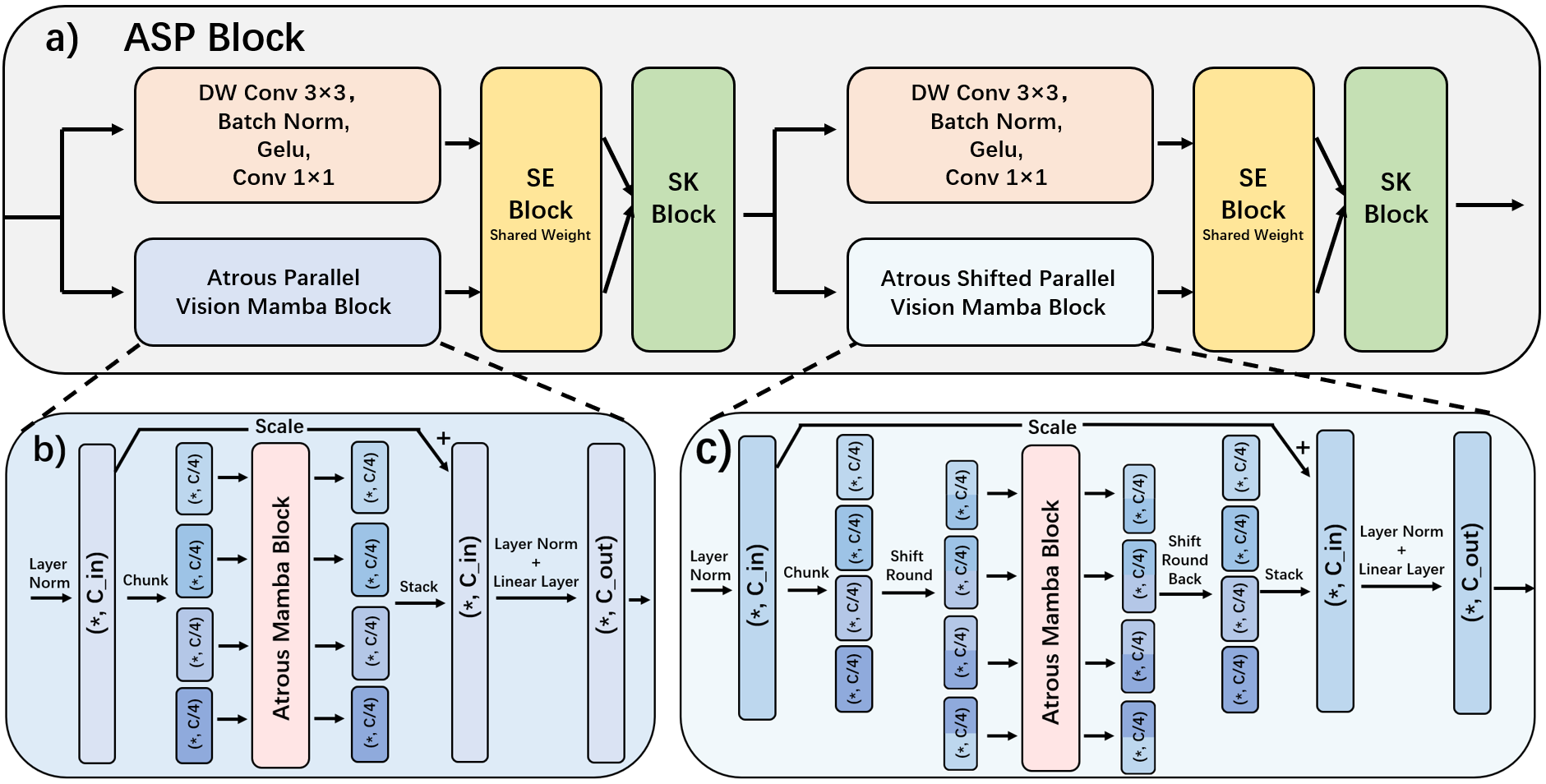}  
    \caption{
    (a) shows the architecture of the Atrous Shifted Parallel (ASP) Block, which consists of a Mamba branch and a CNN branch, followed by a shared-weighted SE block and an SK block. This structure is repeated twice. The first Mamba block is the Atrous Parallel Vision Mamba (APVM) block, as shown in (b), while the second Mamba block is the Atrous Shifted Parallel Vision Mamba (ASPVM) block, incorporating the \textit{shift round} operation, as depicted in (c).
    }
    \label{fig:ASP}
\end{figure*}

Except for stage 1, all the remaining stages use ASP Block, whose structure is shown in Fig. \ref{fig:ASP}a, containing one CNN branch and one Mamba branch. As for the Mamba branch, features pass through the Atrous Parallel Vision Mamba (APVM) Block, which follows Parallel Vision Mamba (PVM) layer\cite{ultralightVMUnet} as the main structure and employs $atrous$ $scan$ as scan strategy shown in Fig. \ref{fig:ASP}b. The details of the APVM Block will be introduced in Section \ref{sub:APVM Block}. Additionally, we introduce $shift$ $round$ operation in $C$-dimension to enhance information exchange and expression on the base of APVM Block. As shown in Fig. \ref{fig:ASP}c, the second Mamba branch is the Atrous Shifted Parallel Vision Mamba (ASPVM) Block, which incorporates $shift$ $round$ operation on the base of APVM Block. The detail of $shift$ $round$ operation and ASPVM block will be introduced in Section \ref{sub:AsPVM Block}.

As for the CNN branch, following \cite{efficientScan}, we design a CNN branch to supplement global features with local features. The architecture of this CNN branch is comprised of a depth-wise CNN with a kernel size of 3, followed by batch normalization, a GELU activation function, and a subsequent CNN with a 1×1 kernel size.
Still following \cite{efficientScan}, we design a shared-weighted Squeeze-Excitation (SE) Block \cite{SENet} to dynamically adjust the weight of the channel according to the feature map. After SE Block, we employ one Selective Kernel (SK) Block \cite{SKNet} to dynamically fusion global feature information and local feature information by assigning different attention to the global and local features.

\subsubsection{Atrous Parallel Vision Mamba Block} \label{sub:APVM Block}
Learning from \cite{ultralightVMUnet}, we propose the APVM Block, as illustrated in Fig. \ref{fig:ASP}b. The key idea is to chunk the feature map along the $C$-dimension into several segments. All feature segments are passed into the same Atrous Mamba Block in parallel, which significantly reduces the number of parameters while maintaining competitive performance. For instance, in this paper, the input features are divided into four segments, leading to a reduction of 74.8\% in the number of parameters compared with the standard Mamba Block. After processing by the Atrous Mamba Block, the segmented features are concatenated along the $C$-dimension to restore the original dimension. Additionally, a learnable adjustment factor is introduced to perform residual connections, enhancing the model's ability to capture long-range spatial dependencies \cite{lightmunet}. Finally, the output is obtained by applying Layer Normalization followed by a linear transformation The process can be formulated as follows:

\begin{align}
    &Y_1^{C/4}, Y_2^{C/4}, Y_3^{C/4}, Y_4^{C/4} = Chunk(LN(Input^{C})) \label{eq:7}\\
    &\overline{Y_i^{C/4}} = Mamba(Y_i^{C/4}), i=1,2,3,4 \label{eq:8}\\
    &O^{C} = Cat([\overline{Y_1^{C/4}},\overline{Y_2^{C/4}},\overline{Y_3^{C/4}},\overline{Y_4^{C/4}]}) + \theta \times Input \label{eq:9}\\
    &Output^{C} = Linear(LN(O^{C})) \label{eq:10}
\end{align}
where $LN$ is Layer Normalization; $Chunk$ is to divide input features in $C$-dimension; $Mamba$ represent Atrous Mamba Block, described in Section \ref{sub:Atrous Scan}; $\theta$ is learnable factor for residual connection.

\subsubsection{Atrous Shifted Parallel Vision Mamba Block} \label{sub:AsPVM Block}

One disadvantage of APVM Block is that there is insufficient information exchange between feature segments. Such inadequacy hinders the interaction and expression of information, potentially damaging the transmission of detailed information. Therefore, inspired by Swin-Transformer \cite{SwinTransformer}, we propose $shift$ $round$ operation to address this problem.

Before all feature segments enter the Atrous Mamba Block, we conduct $shift$ $round$ operation to all feature segments, which can generate new feature segments. 
\textit{shift round} refers that new feature segments are generated by combining half of the adjacent feature segments with each other, imagining that all feature segments form a continuous circle (see different colors of feature segments for clarity in Fig. \ref{fig:ASP}c). This operation can further enhance information exchange since this operation shifts $C$-dimension rather than Height- and Weight-dimension, .
After passing through Atrous Mamba Block, all feature segments need to shift round back to their original locations.
In this way, channel feature information can be interacted within the newly generated feature segments, thus solving the problem of insufficient interaction.
The whole operation can be formulated as follows:

\begin{align}
    &Y_1^{C/8},......, Y_7^{C/8}, Y_8^{C/8} = Chunk(LN(Input^{C})) \label{eq:11}\\
    &SY_1^{C/4},..., SY_4^{C/4} = Cat([Y_2, Y_3]),..., Cat([Y_8, Y_1]) \label{eq:12} \\
    &\overline{SY_i^{C/4}} = Mamba(SY_i), i=1,2,3,4 \label{eq:13}\\
    &O_2^{C/8},..., O_8^{C/8}, O_1^{C/8} = Chunk(Cat[\overline{SY_1},..., \overline{SY_4}]) \label{eq:14}\\
    &\overline{O_1^{C/4}},...,\overline{O_4^{C/4}} = Cat([O_1, O_2]), ..., Cat([O_7, O_8]) \label{eq:15}\\
    &Out^{C} = Cat([\overline{O_1}, \overline{O_2}, \overline{O_3}, \overline{O_4}]) + \theta \times Input \label{eq:16} \\
    &Output^{C} = LN(Linear(Out)) \label{eq:17}
\end{align}
where Eq. \ref{eq:11} and Eq. \ref{eq:12} represent \textit{shift round} operation, and Eq. \ref{eq:14} and Eq. \ref{eq:15} represent \textit{shift round back} operation.

\subsubsection{SE Block} \label{sec:se}
\begin{figure}[ht]  
    \centering
    \includegraphics[width=0.5\textwidth]{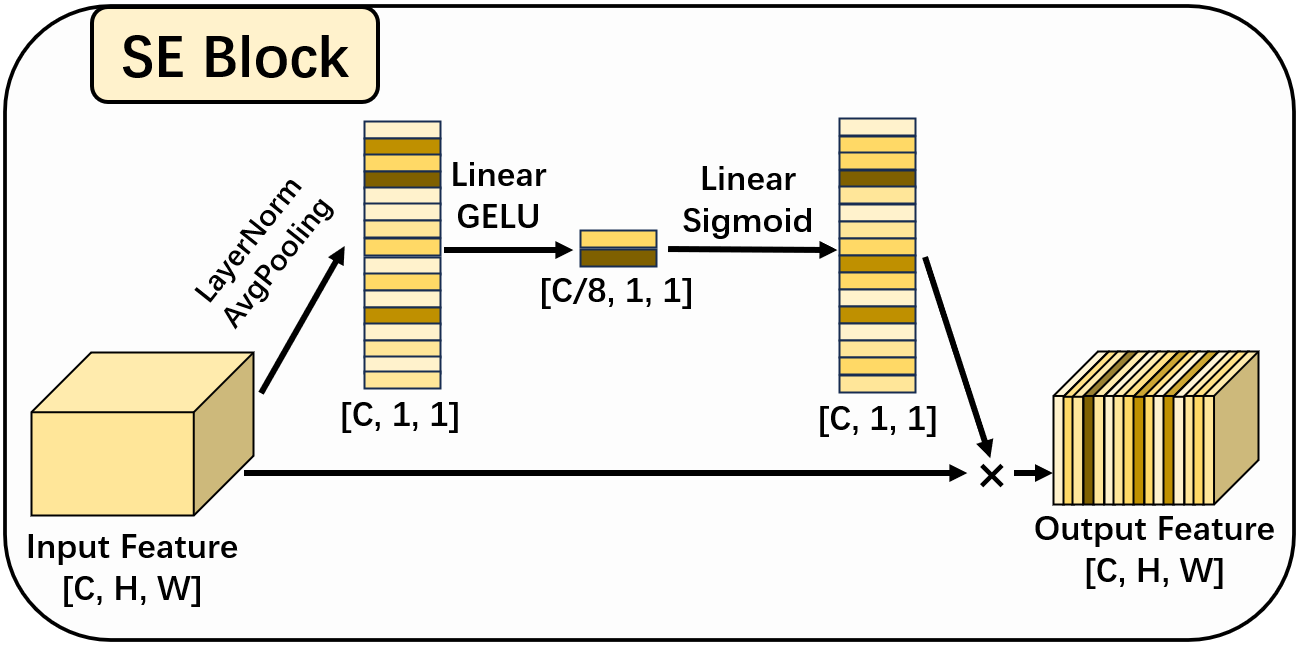}  
    \caption{\centering The architecture of SE Block.}
    \label{fig:SE}
\end{figure}

Following the work of \cite{efficientScan}, a shared-weighted SE Block is designed to be used for both the CNN and Mamba branches. The architecture of the SE Block is illustrated in Fig. \ref{fig:SE}. The preliminary channel attention is derived by LayerNorm and AvgPooling. This channel attention is subsequently squeezed and excited through two distinct linear layers, allowing for the re-adjustment and calibration of the channel attention relationships while simultaneously reducing the number of parameters and FLOPs. Ultimately, the channel attention is multiplied by the input features to produce the output. By incorporating the SE Block, the network is enabled to leverage global information, selectively enhance informative features, and suppress irrelevant features along the $C$-dimension.

\subsubsection{SK Block} \label{sec:sk}
\begin{figure}[ht]  
    \centering
    \includegraphics[width=0.5\textwidth]{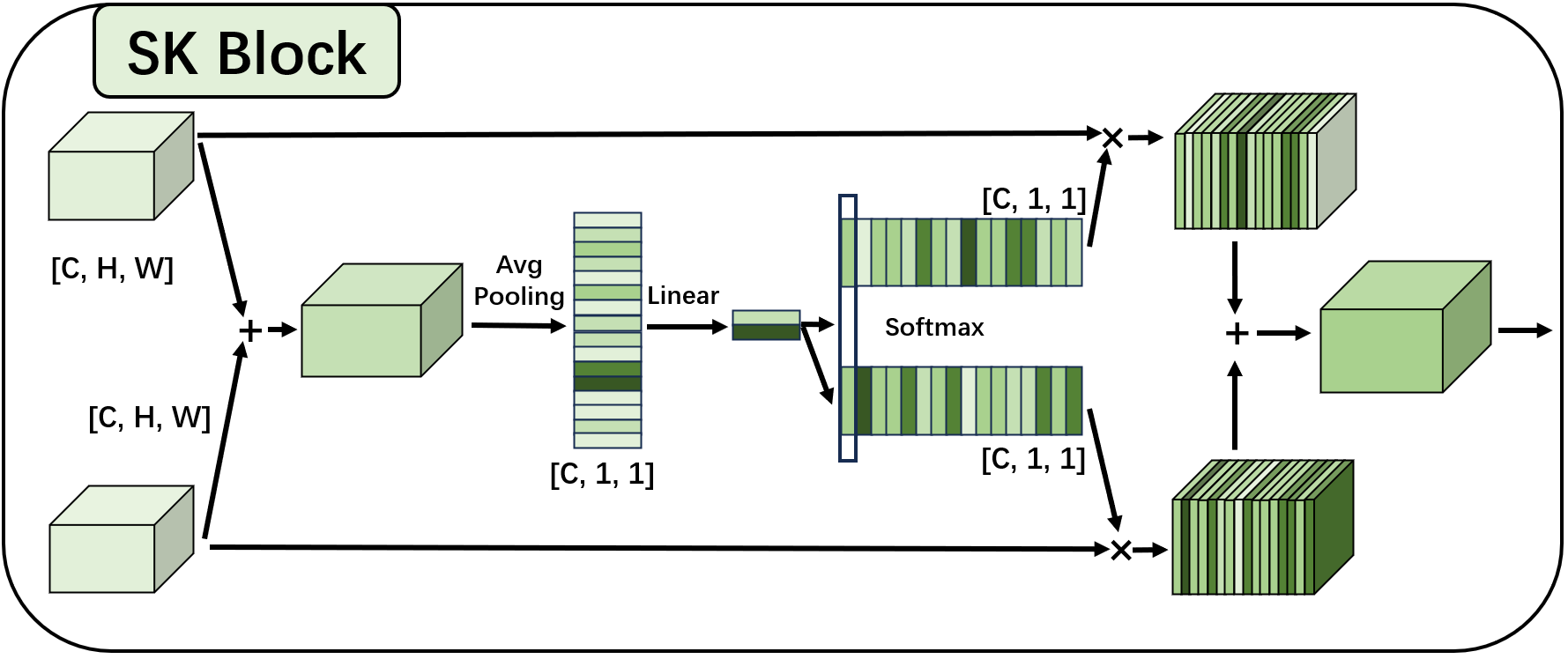}  
    \caption{\centering The architecture of SK Block.}
    \label{fig:SK}
\end{figure}

Some studies \cite{efficientScan, DeformMamba} have also proposed architectures that incorporate one CNN branch and one Mamba branch to capture local and global features. However, these approaches typically employ a direct addition of global and local features, which presents the disadvantage of an incomplete and non-comprehensive information fusion. Inspired by \cite{SKNet}, we employed one SK Block to facilitate the fusion of global and local information. The SK Block merges local and global information, guided by the attention weights generated through the Softmax function. Softmax function assigns attention to local and global information. The fused information is the addition of two features weighted by corresponding weights.
This allows the model to effectively fusion these two types of features. The detailed structure of the SK Block is illustrated in Fig. \ref{fig:SK}.

\subsection{Atrous Mamba Block} \label{sub:Atrous Scan}

\begin{figure}[ht]  
    \centering
    \includegraphics[width=0.5\textwidth]{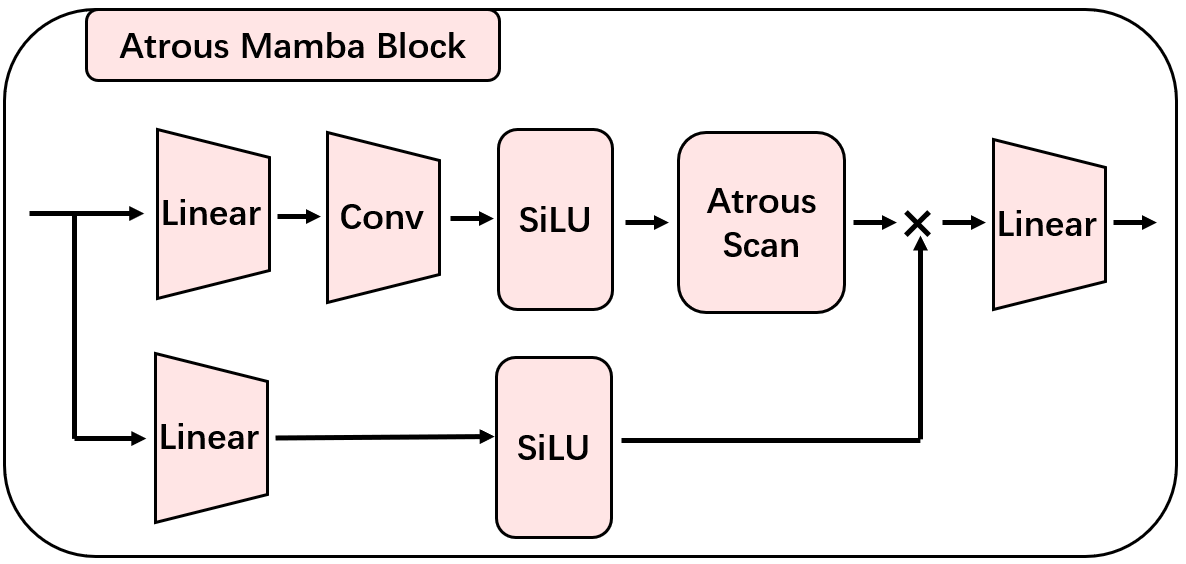}  
    \caption{\centering The architecture of Atrous Mamba Block.}
    \label{fig:AtrousMambaBlock}
\end{figure}

The architecture of the Atrous Mamba Block is illustrated in Fig. \ref{fig:AtrousMambaBlock}. The overall structure aligns with the majority of existing works \cite{ultralightVMUnet, efficientScan, DeformMamba, VMUNet, VMUNetv2}, but utilizes $atrous$ $scan$ as its scanning strategy.

As for images in skin lesions, we find the backgrounds normally occupy large spaces. Therefore, when Mamba executes the scanning, it will inevitably input some less informative patches, so that the hidden state retains some meaningless information, leading to potential performance degradation. Therefore, inspired by atrous CNN \cite{atrousCNN}, we propose $atrous$ $scan$ to address this issue.

The process of $atrous$ $scan$ is illustrated in Fig. \ref{fig:Atrous Scan}. Let $S$ denote the atrous step, and $H$ and $W$ represent the height and width of an image in terms of patches. Initially, to facilitate batch processing, the images are padded such that $H+P_H$ and $W+P_W$ are divisible by $S$, where $P_H$ and $P_W$ are the sizes of padding-bottom and padding-right, respectively. Subsequently, similar to the skipping mechanism in atrous CNNs \cite{atrousCNN}, the images undergo atrous sampling as shown in Fig. \ref{fig:Atrous Scan}. This process generates $S^2$ sub-images. These $S^2$ sub-images are flattened into 1D sequences, each of length $(H+P_H)(W+P_W)/S^2$. These $S^2$ 1D patch sequences are then stacked and passed into the vallian Mamba. Finally, the patches are returned to their original spatial positions, and the padding is removed to restore the images to their initial dimensions.

\begin{figure}[ht]  
    \centering
    \includegraphics[width=0.5\textwidth]{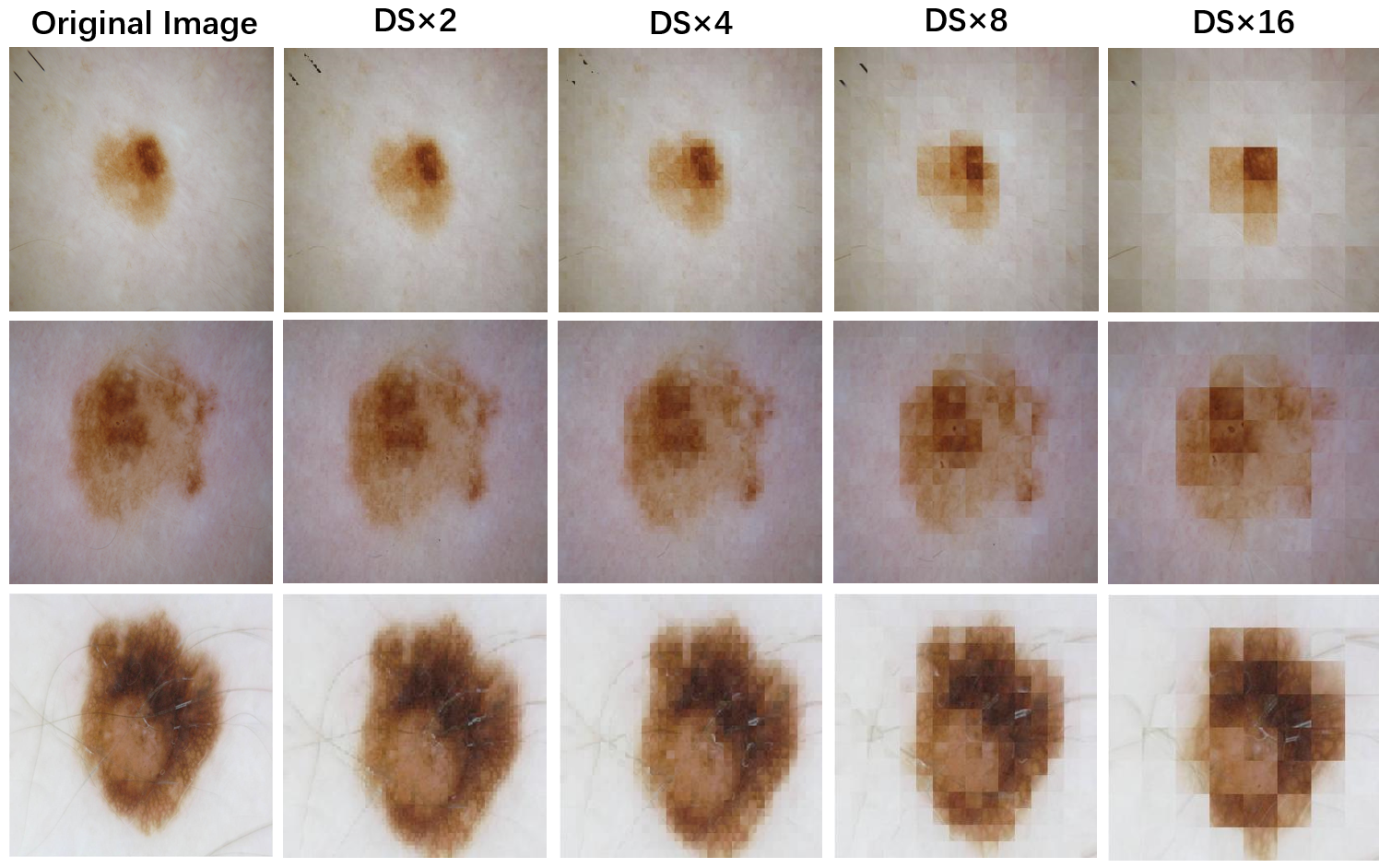}  
    \caption{Three real skin lesion example images that are sampled by \textit{atrous scan} with atrous step equal to 2 at various Down-Sampling rates (DS).}
    \label{fig:SampleExp}
\end{figure}

Fig. \ref{fig:SampleExp} showcases examples of sampled images at four down-sampling rates when atrous step $S$ is set as 2. Three example images in \ref{fig:SampleExp} preserve crucial information, including the shape, color, and texture of the skin lesion, while reducing the large number of meaningless background patches. This differs fundamentally from pooling and strided convolution methods. While these traditional down-sampling techniques diminish resolution and expand the receptive field, they often sacrifice important information or directly discard some patches. In contrast, although $atrous$ $scan$ reduces the resolution by sampling, all patches in different sub-images are subsequently processed by the vallian Mamba. As a result, $atrous$ $scan$ successfully retains all essential information while reducing the impact of irrelevant patches.

\section{Why Atrous Scan is Different?}

\subsection{Scan Strategy Category} \label{sec:category}

\begin{figure*}[ht]  
    \centering
    \includegraphics[width=0.95\textwidth]{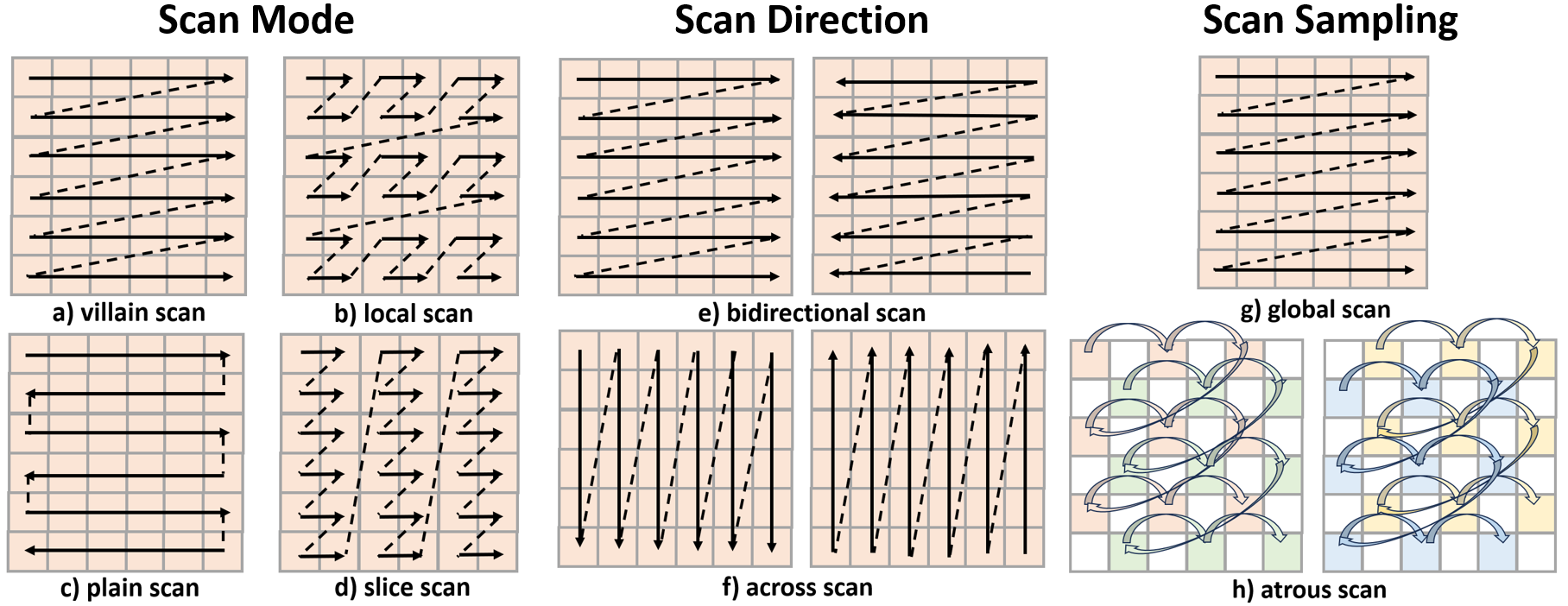}  
    \caption{The scan method can be categorized into three dimensions: scan mode, scan direction, and scan sampling. The figure illustrates several examples of these categories. The scan modes show (a) villain scan, (b) local scan \cite{LocalMamba}, (c) plain scan \cite{plainMamba}, and (d) slice scan \cite{sliceMamba}. The scan directions show (e) bidirectional scan \cite{VisionMamba} and (f) across scan \cite{VMamba}. Across scan includes four sub-figures in Scan Direction. Finally, the scan samplings include (g) global scan and (h) atrous scan.}
    \label{fig:SampleStrategy}
\end{figure*}

There have been numerous scan strategies, such as bidirectional scan \cite{VisionMamba}, across scan \cite{VMamba}, local scan \cite{LocalMamba}, plain scan \cite{plainMamba}, efficient scan \cite{efficientScan} and so on. While one survey \cite{survey1} has divided scan strategies into 4 categories, we want to re-define the categorization of scan strategy to show the uniqueness of \textit{atrous scan}. In this paper, we divide the scan strategies into three categories, i.e. scan mode, scan direction and scan sampling. Fig. \ref{fig:SampleStrategy} illustrates some examples of each category.

\begin{itemize}

    \item \textbf{Scan Mode:} Since Mamba operates as a time-series model, a 2D image has to be transformed into 1D sequences. Therefore, scan mode can be defined as what mode (pattern) it takes to transform a 2D image into a 1D sequence. Different scan modes offer distinct advantages. See Fig. \ref{fig:SampleStrategy} for clarity.

    \item \textbf{Scan Direction:} The transformation of a 2D image into a 1D sequence typically begins with the upper-left patch and initially scans horizontally. However, some studies investigate other scan directions and begin from various starting points, including the top left, top right, bottom left, and bottom right patch, as well as different initial scanning orientations (both horizontal and vertical) to enhance the representation of 2D spatial information. 
    Unlike scan mode, the scan direction is constrained to a total of 8 directions (4 starting corner patches × 2 initial scan directions) in 2D data, and 24 directions in 3D data. 
    Starting points other than four corners (or eight in 3D data) are meaningless.
    
    \item \textbf{Scan Sampling:} Scan sampling is an aspect often overlooked in existing literature, which first samples the image into several sub-images. Existing approaches typically employ a global scan sampling, utilizing the entire image for subsequent scanning, shown in Fig. \ref{fig:SampleStrategy}.g. In contrast, \textit{atrous scan} first samples the image into $S^2$ sub-images. To the best of our knowledge, \textit{atrous scan} is the first scan sampling strategy besides global sampling.

\end{itemize}

In our categorization, it is important to note that a complete scan method requires selecting one method from each of the following categories: scan mode, scan direction, and scan sampling. For instance, the bidirectional scan \cite{VisionMamba} consists of a vallian scan mode, a bidirectional scan direction, and global scan sampling. In this context, the $atrous$ $scan$ can serve as a plug-and-play scan method, which can be integrated into existing scan methods to enhance performance. In a later ablation study, we demonstrated that combining $atrous$ $scan$ with both vallian scan and across scan can improve performance in Section \ref{sec:ablationAtrousScan}.

\subsection{Difference with Efficient Scan } \label{sec:diffWithEfficientScan}
\begin{figure}[ht]  
    \centering
    \includegraphics[width=0.47\textwidth]{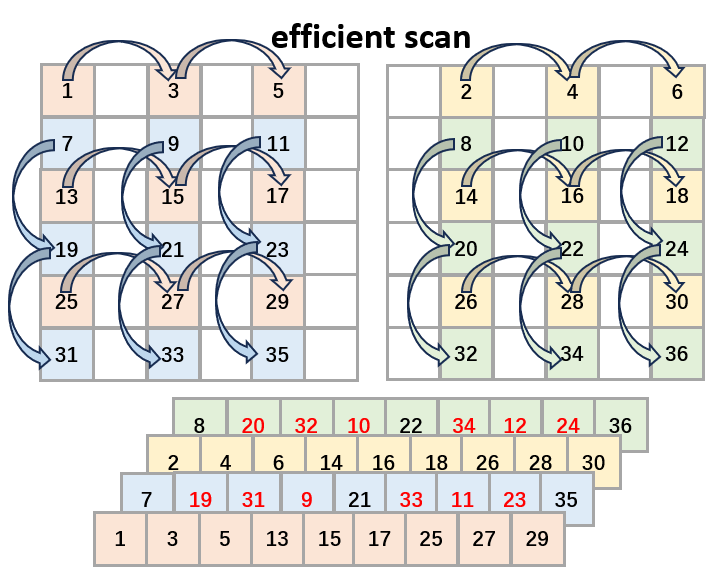}  
    \caption{The diagram of efficient scan \cite{efficientScan}. The four 1D sequences below the image are obtained by efficient scan. The numbers marked in red are those that are different from the atrous scan.}
    \label{fig:efficientScan}
\end{figure}

A close-related work, efficient scan \cite{efficientScan}, also incorporates the concept of atrous, as depicted in Fig. \ref{fig:efficientScan}. However, atrous scan differs from efficient scan in two key aspects. 1) Efficient scan primarily aims to mitigate the substantial computational load associated with cross scans \cite{VMamba}, whereas our work utilizes the concept of atrous to reduce the impact of less informative patches and increase the receptive field, aiming at enhancing overall performance; 2) Efficient scan does not consider atrous steps greater than 2, while our work presents a generalized implementation that supports various atrous steps. 
However, it is unnecessary for the efficient scan to employ different scanning directions for different sub-images. As shown in Fig. \ref{fig:efficientScan}, the orange and yellow patches are scanned horizontally, while the blue and green patches are scanned vertically. This disrupts the spatial information of the images, resulting in performance degradation, as further evidenced by the subsequent ablation study in Section \ref{sec:ablationAtrousScan}.

\section{Experiment Result}

\subsection{Datasets}
\textbf{ISIC16 \cite{ISIC16}, ISIC17 \cite{ISIC17} and ISIC18 \cite{ISIC18}: }The International Skin Imaging Collaboration\footnote{official website: https://challenge.isic-archive.com} 2016, 2017, and 2018, referred to as ISIC16, ISIC17, and ISIC18, respectively, provides three publicly available skin lesion segmentation datasets. This paper follows the official dataset divisions. Specifically, the ISIC16 dataset consists of a training set with 900 images and a test set with 379 images. The ISIC17 dataset is divided into a training set of 2,000 images, a validation set of 150 images, and a test set of 600 images. The ISIC18 dataset includes 2,594 images for training, 100 for validation, and 1,000 for testing.

\textbf{PH2 \cite{ph2}: }The PH2 dataset is the public dataset to offer region-based annotations along with segmentation masks. This dataset comprises 200 dermoscopic images. We divide this dataset into a training set of 140 images and a test set of 60 images.

\subsection{Implementation Details}

\begin{table*}[]
    \centering
    \begin{tabular}{lccccccccccc}
        \hline
        \multicolumn{1}{l|}{Dataset}                & \multicolumn{5}{c|}{PH2}                                                                   & \multicolumn{5}{c|}{ISIC16}        & \multicolumn{1}{l}{} \\ \hline
        \multicolumn{1}{l|}{Model}                  & MIOU $\uparrow$                                                          & DSC $\uparrow$                                                          & Acc $\uparrow$                                                          & Spe $\uparrow$                                                          & \multicolumn{1}{c|}{Sen $\uparrow$}                                                           & MIOU $\uparrow$                                                         & DSC $\uparrow$                                                          & Acc $\uparrow$                                                          & Spe $\uparrow$                                                          & \multicolumn{1}{c|}{Sen $\uparrow$}                                                           & FLOP/Para $\downarrow$           \\ \hline
        \multicolumn{12}{l}{\textbf{Tiny Version Model}}                                                            \\ \hline
        \multicolumn{1}{l|}{MALUNet \cite{malunet}}                & 87.08                                                         & 93.09                                                         & 95.52                                                         & 96.88                                                         & \multicolumn{1}{c|}{92.71}                                                         & 83.59                                                         & 91.06                                                         & 94.98                                                         & \cellcolor[HTML]{C0C0C0}{\color[HTML]{000000} \textbf{96.69}} & \multicolumn{1}{c|}{90.64}                                                         & 0.08/0.18            \\
        \multicolumn{1}{l|}{EGEUNet \cite{EGEUNet}}                & 87.49                                                         & 93.33                                                         & 95.64                                                         & 96.55                                                         & \multicolumn{1}{c|}{93.76}                                                         & 83.77                                                         & 81.17                                                         & 95.03                                                         & 96.61                                                         & \multicolumn{1}{c|}{91.02}                                                         & 0.07/0.05            \\
        \multicolumn{1}{l|}{UltraLight-VMUNet \cite{ultralightVMUnet}}      & \cellcolor[HTML]{EFEFEF}{\color[HTML]{000000} \textbf{87.70}} & \cellcolor[HTML]{EFEFEF}{\color[HTML]{000000} \textbf{93.44}} & \cellcolor[HTML]{EFEFEF}{\color[HTML]{000000} \textbf{95.76}} & \cellcolor[HTML]{EFEFEF}{\color[HTML]{000000} \textbf{97.09}} & \multicolumn{1}{c|}{\cellcolor[HTML]{EFEFEF}{\color[HTML]{000000} \textbf{92.98}}} & \cellcolor[HTML]{EFEFEF}{\color[HTML]{000000} \textbf{84.30}} & \cellcolor[HTML]{EFEFEF}{\color[HTML]{000000} \textbf{91.48}} & \cellcolor[HTML]{EFEFEF}{\color[HTML]{000000} \textbf{95.19}} & 96.60                                                         & \multicolumn{1}{c|}{\cellcolor[HTML]{EFEFEF}{\color[HTML]{000000} \textbf{91.62}}} & 0.06/0.05            \\
        \multicolumn{1}{l|}{ASP-VMUNet-tiny (ours)} & \cellcolor[HTML]{C0C0C0}\textbf{89.37}                        & \cellcolor[HTML]{C0C0C0}\textbf{94.39}                        & \cellcolor[HTML]{C0C0C0}\textbf{96.36}                        & \cellcolor[HTML]{C0C0C0}\textbf{97.51}                        & \multicolumn{1}{c|}{\cellcolor[HTML]{C0C0C0}\textbf{93.98}}                        & \cellcolor[HTML]{C0C0C0}\textbf{84.49}                        & \cellcolor[HTML]{C0C0C0}\textbf{91.60}                        & \cellcolor[HTML]{C0C0C0}\textbf{95.26}                        & \cellcolor[HTML]{EFEFEF}{\color[HTML]{000000} \textbf{96.68}} & \multicolumn{1}{c|}{\cellcolor[HTML]{C0C0C0}\textbf{91.64}}                        & 0.19/0.29            \\ \hline
        \multicolumn{1}{l|}{Improvement}            & {\color[HTML]{3531FF} +1.67}                                  & {\color[HTML]{3531FF} +0.95}                                   & {\color[HTML]{3531FF} +0.60}                                  & {\color[HTML]{3531FF} +0.42}                                  & \multicolumn{1}{c|}{{\color[HTML]{3531FF} +1.00}}                                  & {\color[HTML]{3531FF} +0.19}                                  & {\color[HTML]{3531FF} +0.12}                                  & {\color[HTML]{3531FF} +0.07}                                  & {\color[HTML]{FE0000} -0.01}                                  & \multicolumn{1}{c|}{{\color[HTML]{3531FF} +0.02}}                                  & -                    \\ \hline
        \multicolumn{11}{l}{\textbf{Base Version Model}}                       & \multicolumn{1}{l}{} \\ \hline
        \multicolumn{1}{l|}{C2SDG \cite{C2SDG}}                  & 88.08                                                         & 93.66                                                         & 95.92                                                         & 97.49                                                         & \multicolumn{1}{c|}{92.66}                                                         & 84.82                                                         & 91.78                                                         & 95.38                                                         & 96.88                                                         & \multicolumn{1}{c|}{91.57}                                                         & 7.97/22.01           \\
        \multicolumn{1}{l|}{HMTUNet \cite{HMTUNet}}                & 90.47                                                         & 94.99                                                         & 96.75                                                         & 97.71                                                         & \multicolumn{1}{c|}{\cellcolor[HTML]{EFEFEF}{\color[HTML]{000000} \textbf{94.76}}} & 83.13                                                         & 90.79                                                         & 94.84                                                         & 96.69                                                         & \multicolumn{1}{c|}{90.14}                                                         & 25.06/60.36          \\
        \multicolumn{1}{l|}{HVMUNet \cite{HVMUNet}}                & 89.29                                                         & 94.34                                                         & 96.35                                                         & 97.75                                                         & \multicolumn{1}{c|}{93.47}                                                         & 84.86                                                         & 91.81                                                         & 95.42                                                         & \cellcolor[HTML]{C0C0C0}{\color[HTML]{000000} \textbf{97.13}} & \multicolumn{1}{c|}{91.07}                                                         & 0.74/8.97            \\
        \multicolumn{1}{l|}{VMUNet \cite{VMUNet}}                 & \cellcolor[HTML]{EFEFEF}{\color[HTML]{000000} \textbf{90.49}} & \cellcolor[HTML]{EFEFEF}{\color[HTML]{000000} \textbf{95.01}} & \cellcolor[HTML]{EFEFEF}{\color[HTML]{000000} \textbf{96.76}} & \cellcolor[HTML]{C0C0C0}{\color[HTML]{000000} \textbf{97.78}} & \multicolumn{1}{c|}{94.66}                                                         & 85.40                                                         & 92.13                                                         & 95.52                                                         & 96.53                                                         & \multicolumn{1}{c|}{92.95}                                                         & 5.96/35.90           \\
        \multicolumn{1}{l|}{VMUNet-v2 \cite{VMUNetv2}}              & 88.87                                                         & 94.11                                                         & 96.15                                                         & 96.92                                                         & \multicolumn{1}{c|}{94.54}                                                         & \cellcolor[HTML]{EFEFEF}{\color[HTML]{000000} \textbf{85.91}} & \cellcolor[HTML]{EFEFEF}{\color[HTML]{000000} \textbf{92.42}} & \cellcolor[HTML]{EFEFEF}{\color[HTML]{000000} \textbf{95.70}} & 96.72                                                         & \multicolumn{1}{c|}{\cellcolor[HTML]{EFEFEF}{\color[HTML]{000000} \textbf{93.10}}} & 4.40/22.77           \\
        \multicolumn{1}{l|}{ASP-VMUNet-base (ours)} & \cellcolor[HTML]{C0C0C0}\textbf{91.57}                        & \cellcolor[HTML]{C0C0C0}\textbf{95.60}                        & \cellcolor[HTML]{C0C0C0}\textbf{97.13}                        & \cellcolor[HTML]{EFEFEF}{\color[HTML]{000000} \textbf{97.77}} & \multicolumn{1}{c|}{\cellcolor[HTML]{C0C0C0}\textbf{95.8}}                         & \cellcolor[HTML]{C0C0C0}{\color[HTML]{000000} \textbf{86.93}} & \cellcolor[HTML]{C0C0C0}{\color[HTML]{000000} \textbf{93.01}} & \cellcolor[HTML]{C0C0C0}{\color[HTML]{000000} \textbf{96.01}} & \cellcolor[HTML]{EFEFEF}{\color[HTML]{000000} \textbf{96.77}} & \multicolumn{1}{c|}{\cellcolor[HTML]{C0C0C0}\textbf{94.09}}                        & 2.14/4.69            \\ \hline
        \multicolumn{1}{l|}{Improvement}            & {\color[HTML]{3531FF} +1.08}                                  & {\color[HTML]{3531FF} +0.59}                                  & {\color[HTML]{3531FF} +0.37}                                  & {\color[HTML]{FF0000} -0.01}                                  & \multicolumn{1}{c|}{{\color[HTML]{3531FF} +1.04}}                                  & {\color[HTML]{3531FF} +1.02}                                  & {\color[HTML]{3531FF} +0.59}                                  & {\color[HTML]{3531FF} +0.31}                                  & {\color[HTML]{FF0000} -0.36}                                  & \multicolumn{1}{c|}{{\color[HTML]{3531FF} +0.99}}                                  & -                    \\ \hline \hline
        \multicolumn{1}{l|}{Dataset}                & \multicolumn{5}{c|}{ISIC17}    & \multicolumn{5}{c|}{ISIC18}    & \multicolumn{1}{l}{} \\ \hline
        \multicolumn{1}{l|}{Model}                  & MIOU $\uparrow$                                                         & DSC $\uparrow$                                                          & Acc $\uparrow$                                                         & Spe $\uparrow$                                                          & \multicolumn{1}{c|}{Sen $\uparrow$}                                                           & MIOU $\uparrow$                                                         & DSC $\uparrow$                                                          & Acc $\uparrow$                                                          & Spe $\uparrow$                                                          & \multicolumn{1}{c|}{Sen $\uparrow$}                                                           & FLOP/Para $\downarrow$           \\ \hline
        \multicolumn{12}{l}{\textbf{Tiny Version Model}}   \\ \hline
        \multicolumn{1}{l|}{MALUNet \cite{malunet}}                & 72.34                                                         & 83.95                                                         & 92.57                                                         & 95.74                                                         & \multicolumn{1}{c|}{\cellcolor[HTML]{EFEFEF}{\color[HTML]{000000} \textbf{82.31}}} & 77.72                                                         & 87.46                                                         & 93.01                                                         & 95.30                                                         & \multicolumn{1}{c|}{\cellcolor[HTML]{EFEFEF}{\color[HTML]{000000} \textbf{87.12}}} & 0.08/0.18            \\
        \multicolumn{1}{l|}{EGEUNet \cite{EGEUNet}}                & \cellcolor[HTML]{EFEFEF}{\color[HTML]{000000} \textbf{72.45}} & \cellcolor[HTML]{EFEFEF}{\color[HTML]{000000} \textbf{84.02}} & \cellcolor[HTML]{EFEFEF}{\color[HTML]{000000} \textbf{92.76}} & \cellcolor[HTML]{C0C0C0}{\color[HTML]{000000} \textbf{96.51}} & \multicolumn{1}{c|}{80.63}                                                         & \cellcolor[HTML]{EFEFEF}{\color[HTML]{000000} \textbf{77.96}} & \cellcolor[HTML]{EFEFEF}{\color[HTML]{000000} \textbf{87.61}} & \cellcolor[HTML]{EFEFEF}{\color[HTML]{000000} \textbf{93.12}} & \cellcolor[HTML]{C0C0C0}{\color[HTML]{000000} \textbf{95.55}} & \multicolumn{1}{c|}{86.89}                                                         & 0.07/0.05            \\
        \multicolumn{1}{l|}{UltraLight-VMUNet \cite{ultralightVMUnet}}      & {\color[HTML]{000000} 71.31}                                  & {\color[HTML]{000000} 83.25}                                  & {\color[HTML]{000000} 92.57}                                  & {\color[HTML]{000000} 97.02}                                  & \multicolumn{1}{c|}{{\color[HTML]{000000} 78.19}}                                  & {\color[HTML]{000000} 77.60}                                  & {\color[HTML]{000000} 87.39}                                  & {\color[HTML]{000000} 92.99}                                  & 95.37                                                         & \multicolumn{1}{c|}{{\color[HTML]{000000} 86.84}}                                  & 0.06/0.05            \\
        \multicolumn{1}{l|}{ASP-VMUNet-tiny (ours)} & \cellcolor[HTML]{C0C0C0}\textbf{74.05}                        & \cellcolor[HTML]{C0C0C0}\textbf{85.09}                        & \cellcolor[HTML]{C0C0C0}\textbf{93.08}                        & \cellcolor[HTML]{EFEFEF}{\color[HTML]{000000} \textbf{96.00}} & \multicolumn{1}{c|}{\cellcolor[HTML]{C0C0C0}\textbf{83.65}}                        & \cellcolor[HTML]{C0C0C0}\textbf{78.56}                        & \cellcolor[HTML]{C0C0C0}\textbf{87.99}                        & \cellcolor[HTML]{C0C0C0}\textbf{93.23}                        & \cellcolor[HTML]{EFEFEF}{\color[HTML]{000000} \textbf{95.04}} & \multicolumn{1}{c|}{\cellcolor[HTML]{C0C0C0}\textbf{88.60}}                        & 0.19/0.29            \\ \hline
        \multicolumn{1}{l|}{Improvement}            & {\color[HTML]{3531FF} +1.60}                                  & {\color[HTML]{3531FF} +1.07}                                  & {\color[HTML]{3531FF} +0.32}                                  & {\color[HTML]{FE0000} -0.51}                                  & \multicolumn{1}{c|}{{\color[HTML]{3531FF} +1.34}}                                  & {\color[HTML]{3531FF} +0.60}                                  & {\color[HTML]{3531FF} +0.38}                                  & {\color[HTML]{3531FF} +0.11}                                  & {\color[HTML]{FE0000} -0.51}                                  & \multicolumn{1}{c|}{{\color[HTML]{3531FF} +1.48}}                                  & -                    \\ \hline
        \multicolumn{11}{l}{\textbf{Base Version Model}}    & \multicolumn{1}{l}{} \\ \hline
        \multicolumn{1}{l|}{C2SDG \cite{C2SDG}}                  & 72.79                                                         & 84.25                                                         & 93.00                                                         & \cellcolor[HTML]{EFEFEF}{\color[HTML]{000000} \textbf{97.22}} & \multicolumn{1}{c|}{79.35}                                                         & 77.96                                                         & 87.61                                                         & 93.16                                                         & \cellcolor[HTML]{C0C0C0}{\color[HTML]{000000} \textbf{95.77}} & \multicolumn{1}{c|}{86.45}                                                         & 7.97/22.01           \\
        \multicolumn{1}{l|}{HMTUNet \cite{HMTUNet}}                & 70.19                                                         & 82.48                                                         & 92.40                                                         & \cellcolor[HTML]{C0C0C0}{\color[HTML]{000000} \textbf{97.52}} & \multicolumn{1}{c|}{{\color[HTML]{000000} 75.83}}                                  & 77.25                                                         & 87.16                                                         & 93.09                                                         & 96.69                                                         & \multicolumn{1}{c|}{83.82}                                                         & 25.06/60.36          \\
        \multicolumn{1}{l|}{HVMUNet \cite{HVMUNet}}                & \cellcolor[HTML]{EFEFEF}{\color[HTML]{000000} \textbf{73.72}} & \cellcolor[HTML]{EFEFEF}{\color[HTML]{000000} \textbf{84.87}} & \cellcolor[HTML]{EFEFEF}{\color[HTML]{000000} \textbf{93.22}} & 97.15                                                         & \multicolumn{1}{c|}{80.51}                                                         & 78.93                                                         & 88.23                                                         & 93.32                                                         & 94.86                                                         & \multicolumn{1}{c|}{89.37}                                                         & 0.74/8.97            \\
        \multicolumn{1}{l|}{VMUNet \cite{VMUNet}}                 & {\color[HTML]{000000} 73.64}                                  & {\color[HTML]{000000} 84.82}                                  & {\color[HTML]{000000} 92.95}                                  & 95.90                                                         & \multicolumn{1}{c|}{\cellcolor[HTML]{C0C0C0}{\color[HTML]{000000} \textbf{83.42}}} & \cellcolor[HTML]{EFEFEF}{\color[HTML]{000000} \textbf{80.11}} & \cellcolor[HTML]{EFEFEF}{\color[HTML]{000000} \textbf{88.96}} & \cellcolor[HTML]{EFEFEF}{\color[HTML]{000000} \textbf{93.78}} & \cellcolor[HTML]{EFEFEF}{\color[HTML]{000000} \textbf{95.46}} & \multicolumn{1}{c|}{\cellcolor[HTML]{EFEFEF}{\color[HTML]{000000} \textbf{89.47}}} & 5.96/35.90           \\
        \multicolumn{1}{l|}{VMUNet-v2 \cite{VMUNetv2}}              & 73.58                                                         & 84.79                                                         & 93.04                                                         & 96.44                                                         & \multicolumn{1}{c|}{82.06}                                                         & {\color[HTML]{000000} 78.19}                                  & {\color[HTML]{000000} 87.76}                                  & {\color[HTML]{000000} 93.04}                                  & 94.53                                                         & \multicolumn{1}{c|}{{\color[HTML]{000000} 89.20}}                                  & 4.40/22.77           \\
        \multicolumn{1}{l|}{ASP-VMUNet-base (ours)} & \cellcolor[HTML]{C0C0C0}\textbf{75.08}                        & \cellcolor[HTML]{C0C0C0}{\color[HTML]{000000} \textbf{85.77}} & \cellcolor[HTML]{C0C0C0}\textbf{93.48}                        & {\color[HTML]{000000} 96.64}                                  & \multicolumn{1}{c|}{\cellcolor[HTML]{EFEFEF}{\color[HTML]{000000} \textbf{83.25}}} & \cellcolor[HTML]{C0C0C0}\textbf{80.32}                        & \cellcolor[HTML]{C0C0C0}{\color[HTML]{000000} \textbf{89.09}} & \cellcolor[HTML]{C0C0C0}\textbf{93.83}                        & {\color[HTML]{000000} 95.33}                                  & \multicolumn{1}{c|}{\cellcolor[HTML]{C0C0C0}\textbf{89.97}}                        & 2.14/4.69            \\ \hline
        \multicolumn{1}{l|}{Improvement}            & \multicolumn{1}{l}{{\color[HTML]{3531FF} +1.36}}              & \multicolumn{1}{l}{{\color[HTML]{3531FF} +0.90}}              & \multicolumn{1}{l}{{\color[HTML]{3531FF} +0.26}}              & \multicolumn{1}{l}{{\color[HTML]{FF0000} -0.88}}              & \multicolumn{1}{l|}{{\color[HTML]{FE0000} -0.17}}                                  & \multicolumn{1}{l}{{\color[HTML]{3531FF} +0.21}}              & \multicolumn{1}{l}{{\color[HTML]{3531FF} +0.13}}              & \multicolumn{1}{l}{{\color[HTML]{3531FF} +0.05}}              & \multicolumn{1}{l}{{\color[HTML]{FF0000} -0.44}}              & \multicolumn{1}{l|}{{\color[HTML]{3531FF} +0.50}}                                  & -                    \\ \hline
    \end{tabular}
    \caption{The comparison of five performance metrics across eight leading models, which are divided into two groups: the tiny version models (with fewer than 0.5M parameters) and the base version models (with more than 0.5M parameters). The \colorbox{gray!50}{\textbf{bold}} and \colorbox{gray!20}{\textbf{bold}} highlight the best and second-best performances for each metric, respectively. The \textcolor{blue}{blue} and \textcolor{red}{red} indicate improvements and degradations in our model's performance compared to the best-performing model, respectively.}
    \label{tab:comparsionResult}
\end{table*}

We resize all images on four datasets to 256 × 256 pixels. To mitigate the risk of overfitting, we employ several data augmentation techniques, including normalization, random horizontal flipping, random vertical flipping, and random rotation. Regarding hyperparameter, we set \textit{[N1, N2, N3, N4, N5, N6]} as [1, 1, 1, 1, 3, 1]. We propose two ASP-VMUNet variants, namely ASP-VMUNet-tiny and ASP-VMUNet-base. The $C$-dimension is set to [8, 16, 24, 32, 48, 64] for ASP-VMUNet-tiny, and is set to [32, 64, 96, 128, 256, 384] for ASP-VMUNet-base. The batch sizes are configured as 80 for the tiny version and 24 for the base version. We utilize BceDiceLoss as the loss function. The AdamW optimizer \cite{adamW} is employed with a learning rate of 1e-3 and a weight decay of 1e-2. Additionally, we apply CosineAnnealingLR \cite{consineLR} as the scheduler, with a maximum span of 50 iterations and a minimum learning rate of 1e-5. The number of epochs is set to 300. All experiments were conducted on a Linux system using PyTorch 2.3.1, TorchVision 1.18.1, and CUDA 12.1, executed on a single NVIDIA GeForce RTX 4090.

\begin{table*}[]
\centering
\begin{tabular}{cccccc|cccccc}
\hline
\multicolumn{6}{c|}{Dataset}                  & \multicolumn{2}{c|}{PH2}                                                                       & \multicolumn{2}{c|}{ISIC16}                                             & \multicolumn{2}{l}{}      \\ \hline
\multicolumn{6}{c|}{Component}                & \multicolumn{6}{l}{Performance Metric}                                                                                                                                                               \\ \hline
Shift & SR    & AS    & CNN   & SE    & SK    & MIOU $\uparrow$                               & \multicolumn{1}{c|}{DSC $\uparrow$}                                 & MIOU $\uparrow$         & \multicolumn{1}{c|}{DSC $\uparrow$}                                 & GFLOPs $\downarrow$     & Para(M) $\downarrow$    \\ \hline
      &       &       &       &       &       & 89.26                               & \multicolumn{1}{c|}{94.33}                               & 85.55        & \multicolumn{1}{c|}{92.21}                               & 1.62        & 1.75        \\
\checkmark &       &       &       &       &       & 89.12(\textcolor{red}{-0.14})                        & \multicolumn{1}{c|}{94.25(\textcolor{red}{-0.08})}                        & 86.06(\textcolor{blue}{+0.51}) & \multicolumn{1}{c|}{92.51(\textcolor{blue}{+0.30})}                        & 1.58(\textcolor{blue}{-0.04}) & 1.83(\textcolor{red}{+0.08}) \\
      & \checkmark &       &       &       &       & 89.59(\textcolor{blue}{+0.47})                        & \multicolumn{1}{c|}{94.51(\textcolor{blue}{+0.26})}                        & 86.00(\textcolor{red}{-0.06}) & \multicolumn{1}{c|}{92.47(\textcolor{red}{-0.04})}                        & 1.62(\textcolor{red}{+0.04}) & 1.75(\textcolor{blue}{-0.08}) \\
      & \checkmark & \checkmark &       &       &       & 90.19(\textcolor{blue}{+0.60})                        & \multicolumn{1}{c|}{94.84(\textcolor{blue}{+0.33})}                        & 86.51(\textcolor{blue}{+0.51}) & \multicolumn{1}{c|}{92.77(\textcolor{blue}{+0.30})}                        & 1.62(\textcolor{blue}{+0.00}) & 2.10(\textcolor{red}{+0.35}) \\
      & \checkmark & \checkmark & \checkmark &       &       & 90.54(\textcolor{blue}{+0.35}))                       & \multicolumn{1}{c|}{95.04(\textcolor{blue}{+0.20})}                        & 86.21(\textcolor{red}{-0.30}) & \multicolumn{1}{c|}{92.60(\textcolor{red}{-0.17})}                        & 2.10(\textcolor{red}{+0.48}) & 3.06(\textcolor{red}{+0.96}) \\
      & \checkmark & \checkmark & \checkmark & \checkmark &       & 90.26(\textcolor{red}{-0.28})                        & \multicolumn{1}{c|}{94.88(\textcolor{red}{-0.16})}                        & 86.28(\textcolor{blue}{+0.07}) & \multicolumn{1}{c|}{92.63(\textcolor{blue}{+0.03})}                        & 2.13(\textcolor{red}{+0.03}) & 3.27(\textcolor{red}{+0.21}) \\
      & \checkmark & \checkmark & \checkmark &       & \checkmark & 90.87(\textcolor{blue}{+0.61})                        & \multicolumn{1}{c|}{95.21(\textcolor{blue}{+0.33})}                        & 86.59(\textcolor{blue}{+0.31}) & \multicolumn{1}{c|}{92.81(\textcolor{blue}{+0.18})}                        & 2.10(\textcolor{blue}{-0.03}) & 4.48(\textcolor{red}{+1.21}) \\
      & \checkmark & \checkmark & \checkmark & \checkmark & \checkmark & 91.57(\textcolor{blue}{+0.70})                        & \multicolumn{1}{c|}{{\color[HTML]{000000} 95.60(\textcolor{blue}{+0.39})}} & 86.93(\textcolor{blue}{+0.34}) & \multicolumn{1}{c|}{{\color[HTML]{000000} 93.01(\textcolor{blue}{+0.20})}} & 2.14(\textcolor{red}{+0.04}) & 4.69(\textcolor{red}{+0.21}) \\ \hline
\multicolumn{6}{l|}{Total Improvement}        & \textcolor{blue}{+2.31}                               & \multicolumn{1}{c|}{\textcolor{blue}{+1.27}}                               & \textcolor{blue}{+1.38}        & \multicolumn{1}{c|}{\textcolor{blue}{+0.80}}                               & \textcolor{red}{+0.52}       & \textcolor{red}{+2.94}       \\ \hline \hline
\multicolumn{6}{c|}{Dataset}                  & \multicolumn{2}{c|}{ISIC17}                                                                    & \multicolumn{2}{c|}{ISIC18}                                             & \multicolumn{2}{l}{}      \\ \hline
\multicolumn{6}{c|}{Component}                & \multicolumn{6}{l}{Performance Metric}                                                                                                                                                               \\ \hline
Shift & SR    & AS    & CNN   & SE    & SK    & MIOU $\uparrow$                               & DSC $\uparrow$                                                     & MIOU $\uparrow$        & \multicolumn{1}{c|}{DSC $\uparrow$}                                 & GFLOPs $\downarrow$     & Para(M) $\downarrow$    \\ \hline
      &       &       &       &       &       & 72.48                               & \multicolumn{1}{c|}{84.05}                               & 78.38        & \multicolumn{1}{c|}{87.88}                               & 1.62        & 1.75        \\
\checkmark &       &       &       &       &       & {\color[HTML]{000000} 72.76(\textcolor{blue}{+0.28})} & \multicolumn{1}{c|}{84.23(\textcolor{blue}{+0.18})}                        & 78.88(\textcolor{blue}{+0.50}) & \multicolumn{1}{c|}{88.20(\textcolor{blue}{+0.32})}                        & 1.58(\textcolor{blue}{-0.04}) & 1.83(\textcolor{red}{+0.08}) \\
      & \checkmark &       &       &       &       & 73.25(\textcolor{blue}{+0.49})                        & \multicolumn{1}{c|}{84.56(\textcolor{blue}{+0.33})}                        & 79.14(\textcolor{blue}{+0.26}) & \multicolumn{1}{c|}{88.35(\textcolor{blue}{+0.15})}                        & 1.62(\textcolor{red}{+0.04}) & 1.75(\textcolor{blue}{-0.08}) \\
      & \checkmark & \checkmark &       &       &       & 73.91(\textcolor{blue}{+0.66})                        & \multicolumn{1}{c|}{85.00(\textcolor{blue}{+0.44})}                        & 79.63(\textcolor{blue}{+0.49}) & \multicolumn{1}{c|}{88.66(\textcolor{blue}{+0.31})}                        & 1.62(\textcolor{blue}{+0.00}) & 2.10(\textcolor{red}{+0.35}) \\
      & \checkmark & \checkmark & \checkmark &       &       & 73.03(\textcolor{red}{-0.88})                        & \multicolumn{1}{c|}{84.41(\textcolor{red}{-0.59})}                        & 79.23(\textcolor{red}{-0.40}) & \multicolumn{1}{c|}{88.41(\textcolor{red}{-0.25})}                        & 2.10(\textcolor{red}{+0.48}) & 3.06(\textcolor{red}{+0.96}) \\
      & \checkmark & \checkmark & \checkmark & \checkmark &       & 74.03(\textcolor{blue}{+1.00})                        & \multicolumn{1}{c|}{85.08(\textcolor{blue}{+0.67})}                        & 80.38(\textcolor{blue}{+1.15}) & \multicolumn{1}{c|}{89.12(\textcolor{blue}{+0.71})}                        & 2.13(\textcolor{red}{+0.03}) & 3.27(\textcolor{red}{+0.21}) \\
      & \checkmark & \checkmark & \checkmark &       & \checkmark & 74.68(\textcolor{blue}{+0.65})                        & \multicolumn{1}{c|}{85.50(\textcolor{blue}{+0.42})}                        & 79.93(\textcolor{red}{-0.45}) & \multicolumn{1}{c|}{88.85(\textcolor{red}{-0.27})}                        & 2.10(\textcolor{blue}{-0.03}) & 4.48(\textcolor{red}{+1.21}) \\
      & \checkmark & \checkmark & \checkmark & \checkmark & \checkmark & 75.08(\textcolor{blue}{+0.40})                        & \multicolumn{1}{c|}{85.77(\textcolor{blue}{+0.27})}                        & 80.32(\textcolor{blue}{+0.39}) & \multicolumn{1}{c|}{89.09(\textcolor{blue}{+0.24})}                        & 2.14(\textcolor{red}{+0.04}) & 4.69(\textcolor{red}{+0.21}) \\ \hline
\multicolumn{6}{l|}{Total Improvement}        & \textcolor{blue}{+2.60}                               & \multicolumn{1}{c|}{\textcolor{blue}{+1.72}}                               & \textcolor{blue}{+1.94}        & \multicolumn{1}{c|}{\textcolor{blue}{+1.21}}                               & \textcolor{red}{+0.52}       & \textcolor{red}{+2.94}       \\ \hline
\end{tabular}
\caption{The ablation study of each component. The details of each component are explained in the first paragraph of Section \ref{sec:ablationofComponent}. \checkmark represents using corresponding components. The difference in each row is compared to the data in the previous row. \textcolor{blue}{Blue} and \textcolor{red}{red} represent improvement and degradation, respectively.}
\label{tab:component}
\end{table*}

\subsection{Comparison Result}

We compare our model with eight leading models, including MALUNet \cite{malunet}, EGEUNet \cite{EGEUNet}, UltraLight-VMUNet \cite{ultralightVMUnet}, C2SDG \cite{C2SDG}, HMTUNet \cite{HMTUNet}, HVMUNet \cite{HVMUNet}, VMUNet \cite{VMUNet}, and VMUNet-v2 \cite{VMUNetv2}. We use five performance metrics, including Mean Intersection over Union (MIOU), Dice Similarity Coefficient (DSC), Accuracy (Acc), Sensitivity (Sen), and Specificity (Spe). The performance results are summarized in Table \ref{tab:comparsionResult}.

Overall, our model demonstrates the SOTA performance in both the tiny and base model versions. The ASP-VMUNet-base outperforms the second-best model by 0.59\%, 0.59\%, 0.9\%, and 0.13\% in DSC on the PH2, ISIC16, ISIC17, and ISIC18 datasets, respectively, while ASP-VMUNet-base surpasses the second-best model by 1.08\%, 1.02\%, 1.36\%, and 0.21\% in MIOU, respectively. It should be noticed that ASP-VMUNet increases by 1.39\% in MIOU and 0.86\% in DSC compared with the third-best model, HVMUNet on the ISIC18 dataset, despite slight improvement compared with the second-best model. Other performance metrics exhibit similar trends on both the tiny version and the base version, except for specificity, which shows the degradation.

Regarding the number of parameters and FLOPs, although the ASP-VMUNet-tiny model is larger than other models in the tiny version, the observed performance improvements are obvious, suggesting that the increase in the number of parameters and FLOPs is acceptable. Furthermore, the ASP-VMUNet-base is the lightest among the base-version models, while demonstrating a significant enhancement in performance. In summary, ASP-VMUNet architecture achieves a favorable balance between model performance and computational efficiency.

\subsection{Ablation Study of Component} \label{sec:ablationofComponent}

\begin{table*}[]
\centering
\begin{tabular}{c|ccccc|ccccc|cc}
\hline
Dataset     & \multicolumn{5}{c|}{PH2}                                                                                                                                                                                                          & \multicolumn{5}{c|}{ISIC16}                                                                                                                                                                                & \multicolumn{2}{l}{}  \\ \hline
Atrous Step & MIOU $\uparrow$                                                         & DSC $\uparrow$                                   & Acc $\uparrow$                                   & Spe $\uparrow$                                   & Sen $\uparrow$                                   & MIOU $\uparrow$                                  & DSC $\uparrow$                                   & Acc $\uparrow$                                   & Spe $\uparrow$                                   & Sen $\uparrow$                                   & GFLOPs $\downarrow$ & Parameter(M) $\downarrow$ \\ \hline
1           & 90.96                                                         & 95.27                                  & 96.95                                  & \cellcolor[HTML]{C0C0C0}\textbf{98.21} & 94.34                                  & \cellcolor[HTML]{EFEFEF}\textbf{86.77} & \cellcolor[HTML]{EFEFEF}\textbf{92.92} & \cellcolor[HTML]{EFEFEF}\textbf{95.99} & 96.98                                  & \cellcolor[HTML]{EFEFEF}\textbf{93.44} & 2.136  & 4.35         \\
2           & \cellcolor[HTML]{EFEFEF}{\color[HTML]{000000} \textbf{91.57}} & \cellcolor[HTML]{EFEFEF}\textbf{95.60} & \cellcolor[HTML]{EFEFEF}\textbf{97.13} & 97.77                                  & 95.80                                  & \cellcolor[HTML]{C0C0C0}\textbf{86.93} & \cellcolor[HTML]{C0C0C0}\textbf{93.01} & \cellcolor[HTML]{C0C0C0}\textbf{96.01} & 96.77                                  & \cellcolor[HTML]{C0C0C0}\textbf{94.09} & 2.136  & 4.69         \\
3           & 91.50                                                         & 95.56                                  & 97.08                                  & 97.40                                  & \cellcolor[HTML]{C0C0C0}\textbf{96.43} & 86.50                                  & 92.76                                  & 95.90                                  & 96.92                                  & 93.29                                  & 2.136  & 5.27         \\
4           & 91.03                                                         & 95.30                                  & 96.92                                  & 97.30                                  & \cellcolor[HTML]{EFEFEF}\textbf{96.12} & 86.36                                  & 92.68                                  & 95.87                                  & \cellcolor[HTML]{C0C0C0}\textbf{97.06} & 92.82                                  & 2.136  & 6.08         \\
8           & \cellcolor[HTML]{C0C0C0}\textbf{91.73}                        & \cellcolor[HTML]{C0C0C0}\textbf{95.69} & \cellcolor[HTML]{C0C0C0}\textbf{97.20} & \cellcolor[HTML]{EFEFEF}\textbf{98.01} & 95.51                                  & 86.43                                  & 92.72                                  & 95.88                                  & \cellcolor[HTML]{EFEFEF}\textbf{96.99} & 93.05                                  & 2.136  & 11.64        \\ \hline \hline
Dataset     & \multicolumn{5}{c|}{ISIC17}                                                                                                                                                                                                       & \multicolumn{5}{c|}{ISIC18}                                                                                                                                                                                & \multicolumn{2}{c}{}  \\ \hline
Atrous Step & MIOU $\uparrow$                                                         & DSC $\uparrow$                                   & Acc $\uparrow$                                   & Spe $\uparrow$                                   & Sen $\uparrow$                                   & MIOU $\uparrow$                                  & DSC $\uparrow$                                   & Acc $\uparrow$                                   & Spe $\uparrow$                                   & Sen $\uparrow$                                   & GFLOPs $\downarrow$ & Parameter(M) $\downarrow$\\ \hline
1           & 74.06                                                         & 85.10                                  & 93.16                                  & 96.39                                  & \cellcolor[HTML]{EFEFEF}\textbf{82.71} & 80.11                                  & 88.96                                  & 93.78                                  & 95.45                                  & 89.49                                  & 2.136  & 4.35         \\
2           & \cellcolor[HTML]{EFEFEF}\textbf{75.08}                        & \cellcolor[HTML]{EFEFEF}\textbf{85.77} & 93.48                                  & 96.64                                  & \cellcolor[HTML]{C0C0C0}\textbf{83.25} & \cellcolor[HTML]{C0C0C0}\textbf{80.32} & \cellcolor[HTML]{C0C0C0}\textbf{89.09} & 93.83                                  & 95.33                                  & \cellcolor[HTML]{C0C0C0}\textbf{89.97} & 2.136  & 4.69         \\
3           & 74.56                                                         & 85.43                                  & 93.49                                  & \cellcolor[HTML]{EFEFEF}\textbf{97.40} & 80.84                                  & 80.04                                  & 88.91                                  & 93.73                                  & 95.24                                  & \cellcolor[HTML]{EFEFEF}\textbf{89.85} & 2.136  & 5.27         \\
4           & 74.81                                                         & 85.59                                  & \cellcolor[HTML]{EFEFEF}\textbf{93.52} & 97.21                                  & 81.55                                  & \cellcolor[HTML]{EFEFEF}\textbf{80.19} & \cellcolor[HTML]{EFEFEF}\textbf{89.00} & \cellcolor[HTML]{C0C0C0}\textbf{93.86} & \cellcolor[HTML]{EFEFEF}\textbf{95.82} & 88.80                                  & 2.136  & 6.08         \\
8           & \cellcolor[HTML]{C0C0C0}\textbf{75.63}                        & \cellcolor[HTML]{C0C0C0}\textbf{86.12} & \cellcolor[HTML]{C0C0C0}\textbf{93.77} & \cellcolor[HTML]{C0C0C0}\textbf{97.44} & 81.89                                  & 80.03                                  & 88.91                                  & \cellcolor[HTML]{EFEFEF}\textbf{93.84} & \cellcolor[HTML]{C0C0C0}\textbf{96.02} & 88.23                                  & 2.136  & 11.64        \\ \hline
\end{tabular}
\caption{The ablation study of atrous step. The \colorbox{gray!50}{\textbf{bold}} and \colorbox{gray!20}{\textbf{bold}} highlight the best and second-best performances for each metric, respectively.}
\label{tab:atrous step}
\end{table*}

\begin{table*}[]
\centering
\begin{tabular}{l|ccccccccccc}
\hline
Dataset                                            & \multicolumn{5}{c|}{PH2}                                                                        & \multicolumn{5}{c|}{ISIC16}                                                                     & \multicolumn{1}{l}{} \\ \hline
\multicolumn{1}{c|}{}                              & \multicolumn{11}{l}{Performance Metric}                                                                                                                                                                                  \\ \cline{2-12} 
\multicolumn{1}{c|}{\multirow{-2}{*}{Scan Method}} & MIOU $\uparrow$        & DSC $\uparrow$                                & Acc $\uparrow$  & Spe $\uparrow$  & \multicolumn{1}{c|}{Sen $\uparrow$}   & MIOU $\uparrow$        & DSC $\uparrow$                                & Acc $\uparrow$  & Spe $\uparrow$  & \multicolumn{1}{c|}{Sen $\uparrow$}   & Para $\downarrow$                \\ \hline
Vallian Scan                                       & 90.96        & 95.27                               & 96.95 & 98.21 & \multicolumn{1}{c|}{94.34} & 86.77        & 92.92                               & 95.99 & 96.98 & \multicolumn{1}{c|}{93.44} & 4.35                 \\
+Atrous Sampling                                   & 91.57(\textcolor{blue}{+0.61}) & {\color[HTML]{000000} 95.60(\textcolor{blue}{+0.33})} & 97.13 & 97.77 & \multicolumn{1}{c|}{95.80} & 86.93(\textcolor{blue}{+0.16}) & {\color[HTML]{000000} 93.01(\textcolor{blue}{+0.09})} & 96.01 & 96.77 & \multicolumn{1}{c|}{94.09} & 4.69                 \\ \hline
Across Scan                                        & 91.01        & 95.29                               & 96.94 & 97.77 & \multicolumn{1}{c|}{95.22} & 86.47        & 92.74                               & 95.89 & 96.96 & \multicolumn{1}{c|}{93.16} & 4.69                 \\
+Atrous Sampling                                   & 91.15(\textcolor{blue}{+0.14}) & 95.37(\textcolor{blue}{+0.08})                        & 96.97 & 97.43 & \multicolumn{1}{c|}{96.00} & 86.72(\textcolor{blue}{+0.25}) & 92.89(\textcolor{blue}{+0.15})                        & 95.95 & 96.84 & \multicolumn{1}{c|}{93.69} & 6.08                 \\ \hline
Efficient Scan                                     & 91.01(\textcolor{red}{-0.56}) & 95.29(\textcolor{red}{-0.29})                        & 96.94 & 97.77 & \multicolumn{1}{c|}{95.22} & 86.47(\textcolor{red}{-0.46}) & 92.74(\textcolor{red}{-0.27})                        & 95.89 & 96.96 & \multicolumn{1}{c|}{93.16} & 4.69                 \\ \hline
Dataset                                            & \multicolumn{5}{c|}{ISIC17}                                                                     & \multicolumn{5}{c|}{ISIC18}                                                                     & \multicolumn{1}{l}{} \\ \hline
\multicolumn{1}{c|}{}                              & \multicolumn{11}{l}{Performance Metric}                                                                                                                                                                                  \\ \cline{2-12} 
\multicolumn{1}{c|}{\multirow{-2}{*}{Scan Method}} & MIOU $\uparrow$         & DSC $\uparrow$                                & Acc $\uparrow$  & Spe $\uparrow$  & \multicolumn{1}{c|}{Sen}   & MIOU $\uparrow$        & DSC $\uparrow$                                & Acc $\uparrow$  & Spe $\uparrow$  & \multicolumn{1}{c|}{Sen $\uparrow$}   & Para $\downarrow$                \\ \hline
Vallian Scan                                       & 74.06        & 85.10                               & 93.16 & 96.39 & \multicolumn{1}{c|}{82.71} & 80.11        & 88.96                               & 93.78 & 95.45 & \multicolumn{1}{c|}{89.49} & 4.35                 \\
+Atrous Sampling                                   & 75.08(\textcolor{blue}{+1.02}) & {\color[HTML]{000000} 85.77(\textcolor{blue}{+0.67})} & 93.48 & 96.64 & \multicolumn{1}{c|}{83.25} & 80.32(\textcolor{blue}{+0.21}) & {\color[HTML]{000000} 89.09(\textcolor{blue}{+0.13})} & 93.83 & 95.33 & \multicolumn{1}{c|}{89.97} & 4.69                 \\ \hline
Across Scan                                        & 74.23        & 85.21                               & 93.19 & 96.30 & \multicolumn{1}{c|}{83.13} & 80.07        & 88.93                               & 93.94 & 96.64 & \multicolumn{1}{c|}{87.00} & 4.69                 \\
+Atrous Sampling                                   & 75.65(\textcolor{blue}{+1.42}) & 86.13(\textcolor{blue}{+0.92})                        & 93.76 & 97.35 & \multicolumn{1}{c|}{82.14} & 80.20(\textcolor{blue}{+0.13}) & 89.01(\textcolor{blue}{+0.08})                        & 93.86 & 95.83 & \multicolumn{1}{c|}{88.8}  & 6.08                 \\ \hline
Efficient Scan                                     & 74.31(\textcolor{red}{-0.77}) & 85.26(\textcolor{red}{-0.51})                        & 93.17 & 96.11 & \multicolumn{1}{c|}{83.66} & 79.24(\textcolor{red}{-1.08}) & 88.42(\textcolor{red}{-0.67})                        & 93.26 & 93.78 & \multicolumn{1}{c|}{91.91} & 4.69                 \\ \hline
\end{tabular}
\caption{The ablation study of atrous scan sampling and comparison with efficient scan. The \textcolor{blue}{blue} and \textcolor{red}{red} indicate improvements and degradations in our model's performance compared to the model without atrous scan sampling. Efficient scan is compared with vallian scan with atrous scan sampling.
}
\label{tab:scanSampling}
\end{table*}

We conduct a component ablation study, and results are shown in Tab. \ref{tab:component}. \textit{Shift} represents that, in Fig. \ref{fig:ASP}c, the upper part of most top feature segment (C/8) is not shifted round to combine with the lower part of the most bottom feature segment (C/8) to reform a new feature segment (C/4). We use one additional Atrous Mamba Block to deal with two C/8 feature segments. \textit{SR} represents the $shift$ $round$ operation, and \textit{AS} represents using $atrous$ $scan$. \textit{CNN} represents the CNN branch, including normalization and activation function and two CNN layers. $SE$ and $SK$ represent SE Block and SK Block described in Section \ref{sec:se} and Section \ref{sec:sk}, respectively. We conduct this ablation study on four datasets with the base version and only report MIOU and DSC as performance metrics.

Overall, the components designed in this study collectively contribute to the model's performance enhancement. 
The \textit{Shift} operation improves performance, despite a slight decrease of 0.14\% in MIOU and 0.08\% in DSC on the PH2 dataset. 
The introduction of \textit{shift round} operation further improves the performance across all datasets, except for a minor degradation on ISIC16. This suggests that \textit{shift round} operation can facilitate feature information exchange between four feature segments, thereby improving model performance. 
The \textit{atrous scan} leads to a consistent performance improvement, yielding an average increase of over 0.5\% in MIOU and 0.3\% in DSC on four datasets. This improvement implies that reducing the input of less informative patches and increasing the receptive field allows the model to retain more relevant and significant information in the hidden states, thereby improving overall performance. 
However, the inclusion of a single \textit{CNN} branch results in significant performance degradation across all datasets, except PH2. Notably, on the ISIC17 dataset, MIOU and DSC decreased by 0.88\% and 0.59\%, respectively. This finding suggests that the direct addition is insufficient for effectively fusing global and local features, and even degrades the performance. A more robust fusion block is necessary for the aim that using CNN brach to supplement the global information.
The \textit{SE Block} improves performance across all datasets, except for PH2, demonstrating that the SE block enables adaptive suppression of less important features while enhancing the representation of more salient features, thereby improving model expressiveness. 
The introduction of \textit{SK Block} leads to performance gains on all datasets except ISIC18, indicating that utilizing the attention mechanism helps facilitate the fusion of global and local features, thereby enhancing model performance. Finally, when using SE and SK Block together, the performances are improved on four datasets.

Overall, by incorporating the \textit{shift round} operation, \textit{atrous scan}, CNN branch, SE block, and SK block, the model shows improvements of 2.31\%, 1.38\%, 2.60\%, and 1.94\% in MIOU, and 1.27\%, 0.80\%, 1.72\%, and 1.21\% in DSC across the PH2, ISIC16, ISIC17, and ISIC18 datasets, respectively. These results indicate that all components contribute to the overall performance enhancement, although certain components may slightly degrade performance on specific datasets. Notably, the PH2 dataset consistently deviates from the others, likely due to its smaller size compared to the other datasets.

\subsection{Ablation Study of Atrous Step}

\begin{table*}[]
\centering
\begin{tabular}{c|cccccccccc}
\hline
\multicolumn{1}{c|}{Dataset}       & \multicolumn{2}{c|}{PH2}                                                                                                    & \multicolumn{2}{c|}{ISIC16}                                                                          & \multicolumn{2}{c|}{ISIC17}                                                                          & \multicolumn{2}{c|}{ISIC18}                                                                                                 & \multicolumn{2}{l}{}   \\ \hline
\multicolumn{1}{c|}{Encoder Depth} & \multicolumn{10}{l}{Performance Metric}  \\ \hline
{[}N1, N2, N3, N4, N5, N6{]}       & MIOU $\uparrow$                                  & \multicolumn{1}{c|}{DSC $\uparrow$}                                                           & MIOU $\uparrow$                                  & \multicolumn{1}{c|}{DSC $\uparrow$}                                    & MIOU $\uparrow$                                  & \multicolumn{1}{c|}{DSC $\uparrow$}                                    & MIOU $\uparrow$                                                         & \multicolumn{1}{c|}{DSC $\uparrow$}                                    & GFLOPs $\downarrow$ & Parameters(M) $\downarrow$ \\ \hline
{[}1, 1, 1, 1, 1, 1{]}             & 90.35                                  & \multicolumn{1}{c|}{94.93}                                                         & 86.32                                  & \multicolumn{1}{c|}{92.66}                                  & 74.44                                  & \multicolumn{1}{c|}{85.35}                                  & 79.73                                                         & \multicolumn{1}{c|}{88.72}                                  & 1.88   & 3.50          \\
{[}1, 1, 1, 1, 3, 1{]}             & \cellcolor[HTML]{C0C0C0}\textbf{91.57} & \multicolumn{1}{c|}{\cellcolor[HTML]{C0C0C0}{\color[HTML]{000000} \textbf{95.60}}} & \cellcolor[HTML]{C0C0C0}\textbf{86.93} & \multicolumn{1}{c|}{\cellcolor[HTML]{C0C0C0}\textbf{93.01}} & \cellcolor[HTML]{C0C0C0}\textbf{75.08} & \multicolumn{1}{c|}{\cellcolor[HTML]{C0C0C0}\textbf{85.77}} & \cellcolor[HTML]{C0C0C0}{\color[HTML]{000000} \textbf{80.32}} & \multicolumn{1}{c|}{\cellcolor[HTML]{C0C0C0}\textbf{89.09}} & 2.14   & 4.69          \\
{[}1, 1, 1, 1, 9, 1{]}             & \cellcolor[HTML]{EFEFEF}\textbf{91.10} & \multicolumn{1}{c|}{\cellcolor[HTML]{EFEFEF}\textbf{95.34}}                        & \cellcolor[HTML]{EFEFEF}\textbf{86.85} & \multicolumn{1}{c|}{\cellcolor[HTML]{EFEFEF}\textbf{92.96}} & \cellcolor[HTML]{EFEFEF}\textbf{74.78} & \multicolumn{1}{c|}{\cellcolor[HTML]{EFEFEF}\textbf{85.57}} & \cellcolor[HTML]{EFEFEF}\textbf{79.97}                        & \multicolumn{1}{c|}{\cellcolor[HTML]{EFEFEF}\textbf{88.87}} & 2.90   & 8.28 \\ \hline
\end{tabular}
\caption{The ablation study of the number of encoder layers. The \colorbox{gray!50}{\textbf{bold}} and \colorbox{gray!20}{\textbf{bold}} highlight the best and second-best performances for each metric, respectively.}
\label{tab:architecture}
\end{table*}

In this section, we present an ablation study on the impact of the atrous step $S$ for the atrous scan. Specifically, we conduct experiments where the atrous step takes values of 1, 2, 3, 4, and 8, as summarized in Table \ref{tab:atrous step}. It is important to note that when the atrous step is set to 1, the atrous scan is equivalent to a global scan, which serves as our baseline for comparison.

As shown in Table \ref{tab:atrous step}, as the atrous step increases, the number of parameters also grows. This increase occurs because each sub-image sampled by the atrous scan requires a separate linear layer for projection. Consequently, the number of parameters for these linear layers increases quadratically with the atrous step. For instance, when the atrous step is set to 8, the number of sub-images increases to 64, requiring 64 separate linear layers. This results in a 167\% increase in the number of parameters compared with the global scan. Although this increase in computational complexity is undesirable, the model with the atrous step of 8 can perform best on the PH2 and ISIC17 datasets. This may be because using more separate projection layers can enhance the model's ability to learn more complex representations from the multiple sub-images. In contrast, the model with an atrous step of 2 experiences only a 7\% increase in the number of parameters while consistently improving the performance on four datasets. Therefore, an atrous step of 2 achieves the balance between computational complexity and performance improvement.

\subsection{Ablation Study of Atrous Scan Sampling} \label{sec:ablationAtrousScan}

In this subsection, we investigate the impact of incorporating atrous scan sampling into other scan methods on the performance. Specifically, we introduce atrous scan sampling into two widely-used scanning methods, vallian scan and across scan \cite{VMamba}. Additionally, we also compare the performance with the efficient scan \cite{efficientScan}. We only replace the scan methods of ASP-VMUNet-base. The experiments are conducted on four different datasets as shown in Table \ref{tab:scanSampling}.

When incorporating the atrous scan into both the vallian scan and the across scan, we can find that there are improvements in performance on all four datasets, based on both the vallian scan and the across scan. Notably, on the ISIC17 dataset, the MIOU increases by 1.02\% and 1.42\%, while DSC increases by 0.67\% and 0.92\% for the vallian scan and across scan, respectively. These improvements underscore the effectiveness of the atrous scan sampling in enhancing model performance. However, except for ISIC17, the improvement based on across scan is slight (less than 0.2\%). We argue that this may be caused by the poor fusion between four images starting from four directions, which is the inherent limitation of the across scan.  
Furthermore, when comparing our method with the efficient scan, we find that the efficient scan performs worse on all datasets. As discussed in Section \ref{sec:diffWithEfficientScan}, the efficient scan disrupts the spatial relationships between 2D patches by using two different scan directions in each sub-image, which may hinder the model's ability to effectively capture contextual information. On the other hand, the atrous scan maintains the 2D spatial information by keeping the scan direction the same, which allows the model to better preserve contextual and spatial information and improve performance.

\subsection{Ablation Study of The Number of Encoder Layers}

In this subsection, we conduct experiments to investigate the impact of the number of encoder layers on model performance. The results are presented in Table \ref{tab:architecture}. Since UNet-like architectures typically retain lighter structures in the earlier stages, we focus on modifying the number of layers in stage 5 of the encoder. Specifically, we experiment with three different configurations for stage 5, setting $N5$ as 1, 3, and 9.

From Table \ref{tab:architecture}, we observe that the model with the encoder configuration of [1, 1, 1, 1, 3, 1] achieves the best performance. The model with the encoder configuration of [1, 1, 1, 1, 9, 1] performs slightly worse but still ranks as the second-best. However, setting $N5$ as 9 results in a significant increase in the number of parameters by 136.6\% compared with when $N5$ is set to 1. 
On the other hand, setting $N5$ to 3 only increases 34\% in the number of parameters compared with when $N5$ is set to 1, which is more acceptable. Therefore, our model achieves an optimal trade-off between performance improvement and computational efficiency.

\section{Conclusion}

In this paper, we propose the \textbf{A}trous \textbf{S}hifted \textbf{P}arallel \textbf{V}ision \textbf{M}amba \textbf{UNet} (\textbf{ASP-VMUNet}), which introduces several innovations to improve medical image segmentation. We present the \textit{atrous scan} method, which increases the receptive field while retaining all patches, mitigating the impact of less informative ones. Additionally, the $shift$ $round$ operation enhances feature interaction within the Mamba block, and the SK Block fuses local and global information more effectively. Experimental results show that our model achieves state-of-the-art performance on ISIC16/17/18 and PH2 datasets. We also demonstrate that the atrous scan, as a scan sampling method, can be combined with other techniques for further performance gains. The ablation study highlights the contribution of each component and validates the effectiveness of our design choices.

\section*{Acknowledgment}
This work was supported by the Capital's Funds for Health Improvement and Research (Grant Number: CFH 2024-1-2084).

\bibliography{main}
\bibliographystyle{IEEEtran}

\end{document}